\documentclass[pre,preprintnumbers,amsmath,amssymb,floatfix, 12pt]{revtex4}
\usepackage{graphics}
\usepackage{amsmath}
\usepackage{graphicx,color}
\newcommand \be  {\begin{equation}}
\newcommand \beq {\begin{equation}}
\newcommand \bea {\begin{eqnarray} \nonumber }
\newcommand \ee  {\end{equation}}
\newcommand \eeq {\end{equation}}
\newcommand \eea {\end{eqnarray}}
\newcommand{\beqa}{\begin{eqnarray}}
\newcommand{\eeqa}{\end{eqnarray}}

\newcommand{\nn}{\nonumber\\}

\newcommand{\ket}[1]{|{#1}\rangle}
\newcommand{\bra}[1]{\langle{#1}|}
\newcommand{\vidk}{\ket{-}}
\newcommand{\vidb}{\bra{-}}
\newcommand{\brak}[2]{\langle{#1}|{#2}\rangle}
\newcommand{\braok}[3]{\langle{#1}|{#2}|{#3}\rangle}

\newcommand{\bol}[1]{{\boldsymbol{#1}}}
\begin{document}
\title{ Metastable states, transitions, basins and borders at finite temperatures
}  
\author{Sorin T\u{a}nase-Nicola and Jorge Kurchan}
\affiliation{ 
 PMMH UMR 7636 CNRS-ESPCI,\\ 10, Rue Vauquelin, 75231 Paris CEDEX 05,  France}
\date{\today}


\begin{abstract}
Langevin/Fokker-Planck processes can be immersed in a larger frame by adding
fictitious fermion variables. The (super)symmetry of this larger structure has
 been used to derive 
Morse theory in an  elegant way. The original
physical diffusive motion is retained in the zero-fermion subspace. Here we 
study the subspaces with non-zero fermion number which  yield deep 
information, as well as new computational strategies,
 for barriers, reaction paths, and unstable states -- even
in non-zero temperature situations and when the barriers are of entropic or 
collective nature, as in the
 thermodynamic limit. 
The presentation is self-contained.
\end{abstract}
\maketitle

\section{Introduction}
\label{sec:intro}

Many systems have a dynamics with processes which take place on
distinct timescales. 
The most familiar example is the diffusion in a many-valley energy $E(\bol{x})$  landscape 
at low temperatures:
\begin{equation}
\dot x_i = -\frac{\partial E}{\partial x_i} +\sqrt{2T} \; \eta_i,
\label{Lang}
\eeq
 which consists of  rapid gradient descents into local minima,
and slow `activated' transitions between minima induced by the thermal noise
 $\eta_i$ (Gaussian independent white noises of unit variance).
 The relevant parameter is the inverse temperature $\beta=1/T$: the larger $\beta$ the 
more pronounced the gap between fast intra-valley relaxations and slow activation 
processes $\ln(t_{slow}) \sim \beta \Delta E$  \cite{Hanggi90}.
Another example is that of cooperative systems at finite temperature. 
Consider for example a $d$-dimensional ferromagnet in a magnetic field $h$ pointing 
upward: the state with negative magnetization becomes unstable,  its decay taking a 
time $\ln (t_{slow})~\propto~h^{-(d-1)}$,  the parameter controlling 
the timescale separation is the inverse of the field. In the absence of field, 
 the slow relaxation takes a time $\ln (t_{slow}) \propto  L^{d-1}$, and the 
control  parameter is the size $L$\cite{Langer67}.
From a conceptual point of view, it is important in these situations to characterize the
relevant structures: metastable states and their basins of attraction,
 the reaction paths joining them, and the timescales involved. On the other hand,
in order to efficiently model realistic situations,
one needs to be able to treat the rare `activated' passages in a specific way.
   
Quite generally, a separation between timescales  implies the existence of metastable states,
defined as probability distributions corresponding to situations in which everything
fast has happened and everything slow has not taken place.
If there are  more than two timescales, for example $t_{fast} \ll t_{interm} \ll t_{slow}$
one has metastable states at    $t_{interm}$, and a different set of metastable states at
 $t_{slow}$, the latter
resulting from the fusion or the decay of states defined for $t_{interm}$.
Given how natural the concept of metastability is, it may come as a surprise
that only recently has a  construction of metastable states 
based on the stochastic dynamics been fully established \cite{endnote1}.

The idea is simple: the probability associated with (\ref{Lang}) evolves 
through \cite{Risken96}:
\begin{eqnarray}
\frac{dP(\bol{x},t)}{dt} &=& -H_{FP}\; P(\bol{x},t), \nonumber\\
H_{FP} &= &-\sum_i \frac{\partial}{\partial x_i} \left(T\frac{\partial}{\partial x_i}+
E_{,i}\right),
\label{fokker}
\end{eqnarray} 
(from here onward
we denote derivatives as
 $A_{,i}\equiv \frac{\partial A}{\partial x_i}$ 
and $A_{,ij}\equiv \frac{\partial^2 A}{\partial x_i \partial x_j}$)
where  $H_{FP}$ is the Fokker-Planck operator.
It turns out \cite{endnote1} that if there is a separation of timescales 
$t_{fast}\ll t_{slow}$ in the system,  the spectrum of $H_{FP}$ has a gap: there are 
(say) $K$ eigenvalues of the order
of $ t_{slow}^{-1}$ and all other eigenvalues are at least of the order of $ t_{fast}^{-1}$.
Furthermore, one can show that,  in the limit of large timescale separation, irrespective of its
origin, one can construct exactly $K$ probability distributions corresponding to distinct 
metastable states by linear combinations of the $K$ eigenstates `below the gap'.
The low temperature example is particularly clear: in that case these $K$ 
distributions
are Gaussians of width $\sqrt{T}$ sitting at each of the $K$ local minima.

In a situation with metastability, it becomes interesting and in practice necessary
 to evaluate the time of decay, as well as the spatial probability distribution
of the escape current (i.e. the reaction paths). This may involve identifying the
barrier or `bottleneck' responsible for the slowness of decay.
In the low temperature example, the reaction paths are simply  gradient lines 
connecting two energy minima through a saddle with one unstable direction (of index one). 
The bottlenecks are these saddle-points, and there is a 
considerable variety of methods for their location in high dimensional space \cite{Wales03}.
 
Now,  there is a construction that naturally incorporates
saddle points, and that has been successfully used to derive the relations between
the numbers of saddle points of a function (the energy here) and the topological 
properties of the manifold on which it is defined. These relations are the 
so-called Morse inequalities \cite{Milnor63}, and have been rederived
 in an elegant and elementary way by Witten~\cite{Witten82}.
The construction is a `completion' of the diffusive problem as follows. 
First express the Fokker-Planck
operator in a basis in which it is manifestly Hermitian \cite{Risken96}:
\begin{equation}
H^h_{FP} = e^{\beta E/2} H_{FP}  e^{-\beta E/2} = 
\frac{1}{T}\sum_i \left[- T^2 \frac{\partial^2}{\partial x_i^2}
 + \frac{1}{4} E_{,i}^2 - \frac{T}{2} E_{,ii}\right].
\label{hermi}
\end{equation}
Second, `complicate' the operator and the space by introducing $N$ fermions $a^\dag_i$,
$a_i$, with $i=1,\dots,N$, and:
\beq
H^h = H_{FP}^h +  E_{,ij} a^\dag_j a_i =\frac{1}{T}\sum_i \left[
-T^2 \frac{\partial^2}{\partial x_i^2} + \frac{1}{4} E_{,i}^2 - \frac{T}{2} E_{,ii}\right] + \sum_{ij} E_{,ij} a^\dag_j a_i = (H^h)^\dag.
\label{hhsusy}
\eeq
Within zero-fermion subspace, $H^h$ is just the Hermitian form of the
 Fokker-Planck operator,
and, as we have remarked, for low temperatures
its  eigenstates `below the gap' are related to local
minima. 

The wavefunctions  with one or more fermions are so far  a spurious addition.
Their interest stems from the fact that one can show that
there is a gap in all the spectra associated with any
fermion number, and  the states `below the gap' having
one fermion are  for low temperatures peaked  on
 saddles with one unstable direction, those having two fermions
on saddles with two unstable directions, and in general those with $p$ fermions on saddles
of index $p$.
Using this 
fact and the symmetries of $H^h$, it is then easy to derive Morse inequalities \cite{Witten82} 
-- we shall
review this in Section \ref{sec:morse}.

The  basis that makes $H_{FP}$ Hermitian as in (\ref{hermi}) (we shall in what follow refer 
to it as the `Hermitian basis')
offers the direct way to get to Morse theory, and is the one most often used in the field 
theory literature.  Here, we are mainly interested in the diffusive 
interpretation, at least of the  original zero-fermion subspace, and for this
we must go back to the original basis \cite{endnote2}. The change from the Hermitian to 
the original
basis is made by a multiplication by $e^{\beta E/2}$, and is quite tricky
since it is exponentially large in the relevant parameter --- 
$\beta$ if we are interested in the low temperature limit, 
or the system size for macroscopic systems.
 Because of this reason, 
one has to be very careful because negligible, large deviations in one basis become
of $O(1)$ in the other. Indeed, the strategy we shall follow in this paper
is to rederive the limit wavefunctions in each basis from scratch.

A first question we may ask is how do the eigenstates `below the gap' with fermion
number larger than zero look, for low temperatures,
in the original   basis in which
\begin{equation}
H = H_{FP} + \sum_{ij} E_{,ij} a^\dag_j a_i  = e^{-\beta E/2} H^h e^{\beta E/2}.
\label{susy11}
\end{equation}
 The outcome, as we shall see, is a pleasant surprise: for example one
fermion (right) eigenstates `below the gap' are concentrated not on the saddle, but along
 a narrow (width $\sim \sqrt{T}$) tube following the
gradient line joining minima and passing through the saddle -- the reaction path.
Higher fermion number subspaces (and {\em left} eigenstates) also encode interesting
 information.

The construction yielding Morse theory relies on the low-temperature limit, in which
functions peak on the appropriate structures. Low temperatures are just one instance
of evolution with  widely separated scales. One is naturally led to ask
what happens with the construction we have described in the presence of a timescale
separation generated by some other (collective, entropic...) mechanism.
Again, the answer is pleasant: for example the one fermion eigenstates `below the gap' of
(\ref{susy11}) yield the reaction current distributions between metastable states 
(the latter defined dynamically as outlined after Eq.  (\ref{fokker})).
This generalization will give  us practical strategies for the evaluation
of reaction paths, valid whenever there is timescale separation. From a more
abstract point of view, it will yield 
a precise definition of
   `free energy barrier' in a natural way, without having to rely
on mean-field or any other approximation.

Let us write, for a generic wavefunction $\ket{\bol{\psi}}$, an evolution equation:
\begin{equation}
\frac{d \ket{\bol{\psi}}}{dt} = -H \; \ket{\bol{\psi}}. 
\label{evol11}
\end{equation}
Specializing ${\bol{\psi}}$ to zero fermions we recover the Fokker-Planck 
equation (\ref{fokker}). 
Consider now (\ref{evol11})  but within the one-fermion subspace, in which
functions are of the form 
$\ket{{\bol{\xi}^R}}=\int d^N\!\! x\;\sum_c R_c(\bol{x})\ket{\bol{x}} a^\dag_c \vidk$, with
$\ket{\bol{x}}$  the  basis in  space and $\vidk$  the fermion vacuum. 
It amounts to  an evolution for 
 a {\em vector} function $\bol{R}(\bol{x},t)=(R_1(\bol{x},t),..,R_N(\bol{x},t))$:
\begin{equation}
\frac{dR_c(\bol{x},t)}{dt} = -H_{FP}\; R_c(\bol{x},t) - \sum_b 
\frac{\partial^2 E}{\partial x_c \partial x_b }
 R_b(\bol{x},t).
\label{onef}
\end{equation}
{\em Equation (\ref{onef}) is one of the main instruments of this paper}. 
It evolves a vector field $\bol{R}(\bol{x},t)$
so that it rapidly becomes a linear combination of one fermion states `below the gap'.

In a system with metastable states,
starting from an initial condition, the probability distribution 
$P(\bol{x},t)$ evolves rapidly to a quasi-stationary distribution
corresponding to quasi-equilibrium within one or more metastable
 states. At longer timescales, $P(\bol{x},t)$ will be gradually concentrated
on new, more stable metastable states.
If we  choose the initial condition to be close to  a state,
we get a quick thermalization within such a state.
What we have been discussing up to now suggests that  equation (\ref{onef}) 
does for reaction currents
what the Fokker-Planck equation does for states: depending on the initial conditions, 
$\bol{R}(\bol{x},t)$ tends rapidly to a  reaction current 
between   metastable states, which, in the particular case of very low temperatures,
is a single reaction path. As time passes, other new
current distributions start contributing to $\bol{R}(\bol{x},t)$.

The main point is that
 while escape from a state takes by assumption long times for the probability
distribution $P(\bol{x},t)$, convergence to a reaction current is  for 
$\bol{R}(\bol{x},t)$ immediate (or more precisely,
of the same order of the time it takes for $P(\bol{x},t)$ to stabilize in the closest state).
 Now, a whole set of practical methods, such as simulated annealing and transition
 path sampling,
can be seen as 
ways of implementing  the Fokker-Planck equation ---
the former in a diffusive  and the latter in a functional way.
The same can be done with equation (\ref{onef}): we shall give in this paper
a diffusion equation which is to (\ref{onef})  what the Langevin process is to
(\ref{fokker}), and a path-sampling procedure based on (\ref{onef}) which has the peculiarity
that the paths pile up on the barriers.

To conclude this rather technical introduction, let us summarize what we do in this paper.
We first (section \ref{sec:susy}) introduce the supersymmetric construction in detail, 
stressing in particular
the relation between the original Fokker-Planck basis and the Hermitian basis.
As a first example, in section \ref{sec:morse}  we rederive Morse theory for the case of 
 smooth potentials in a simply connected space.

In section \ref{sec:states}, after briefly reviewing the dynamic definition of metastable states,
we analyze in detail the structure of the one fermion subspace, showing that it 
contains the reaction paths
and also `loops': reactions leading from a state to itself. Although for clarity
we always keep the low-temperature
case in mind, the developments are valid whatever the origin of the timescale separation. 

In section \ref{sec:thebig} we summarize the structure in all fermion subspaces. For the low-temperature
case we discuss the form of all  eigenstates `below the gap'. The derivation of these
results can be made in an explicit way using a diffusive
 dynamics we introduce in section \ref{sec:diffusion}, 
which plays  for the higher fermion number subspaces the role that Langevin equation plays
for the zero-fermion Fokker-Planck evolution.  

In section \ref{sec:path} we construct  a path-sampling process to
locate reaction paths, in which the trajectories are weighed with the usual Langevin 
action plus a Lyapunov exponent associated to each  trajectory.

\section{Supersymmetric quantum mechanics and Fokker-Planck equation}
\label{sec:susy}
 We assume that $E(\bol{x})$ is 
two times differentiable and bounded so that the Gibbs measure exists:
\beq
\int_{R^N} d^Nx\,e^{-\beta E} < \infty.
\label{integre}
\eeq
The dynamics of the system is  given by the Langevin equation (\ref{Lang}).
The probability distribution will evolve according to the Fokker Planck equation
(\ref{fokker}), which  can be seen as a continuity equation for the current \cite{Risken96}:
\begin{equation}
J_i(\bol{x},t) \equiv \left(T\frac{\partial}{\partial x_i}+E_{,i}\right) P(\bol{x},t).
\label{current}
\end{equation}
 We introduce at this point $N$ fermion
creation and annihilation operators $a^\dag_i$ and $a_i$, with
anticommutation relations $[a_i,a^\dag_j]_+=\delta_{ij}$; the fermion
number operator is $N_f=\sum_i a^\dag_ia_i$.
We denote states in the coordinate space as $\ket{\psi}$ and
$\psi(\bol{x}) \equiv \brak{\bol{x}}{\psi}$, using the Dirac bra-ket notation
 of Quantum Mechanics; the  zero-fermion state is $\vidk$, and we denote
states in the product space  (coordinate $\otimes$ fermions) with 
boldface. We say that $\ket{\bol{\psi}}$ has $n$
fermions if:
\begin{equation}
N_f\ket{\bol{\psi}} = n \ket{\bol{\psi}}.
\end{equation}

We define the `charges'
\begin{equation}
 {\bar Q}=-{\mathrm i} \sum_i (T\frac{\partial}{\partial x_i}+ E_{,i})a^\dag_i \;\;\; ; \;\;\; Q=
-{\mathrm i}T \sum_i \frac{\partial}{\partial x_i}a_i,
\label{char}
\end{equation}
which satisfy:
\begin{equation}
 {\bar Q}^2=Q^2=0.
\label{nil}
\end{equation}
We can write the operator
\beq
H = \frac{1}{T} ({\bar Q}+Q)^2=\frac{1}{T} [{\bar Q},Q]_+ = H_{FP} +\sum_{ij}  E_{,ij} a^\dag_j a_i,
\label{hsusy}
\eeq
where
\beq
[H,Q]=[H, {\bar Q}]=0.
\eeq
 $H$  commutes with the fermion number operator 
$N_f$, so that eigenstates are classified according
to their fermion number. Within the zero-fermion space, $H$ is the original 
Fokker-Planck operator. 
These relations are true only for the Fokker-Planck equation 
with  the drift forces  at least locally a gradient.
 $Q$ and $\bar Q$  commute with $H$ and 
 transform  states with an even number
of fermions (bosonic states) into those with an odd number
of fermions (fermionic states) and vice-versa, hence the name `supersymmetry'.
What we have done up to now can be seen as completing the square, and making symmetries
underlying the Fokker-Planck equation {\em with gradient forces} explicit.

One can now make a change of basis such that the charges become Hermitian
 conjugates of one another:
\begin{eqnarray}
Q^h &=& e^{\beta E/2}  Q  e^{-\beta E/2} = -{\mathrm i}
\sum_i ( T\frac{\partial}{\partial x_i} -\frac{1}{2}E_{,i})a_i,  \nn
{\bar Q}^h &=& e^{\beta E/2} {\bar Q} e^{-\beta E/2}  =
-{\mathrm i}\sum_i (T\frac{\partial}{\partial x_i}+\frac{1}{2}E_{,i})a^\dag_i =  (Q^h)^\dag.
 \label{hermiticity}
\eea
 In the new basis we have the Hermitian equivalent
of (\ref{hsusy}):
\bea
H^h &=&\frac{1}{T} [{\bar Q}^h,Q^h]_+ 
=\frac{1}{T} ({\bar Q}^h+Q^h)^2= H_{FP}^h +  \sum_{ij}E_{,ij} a^\dag_j a_i =
\nn
& &\frac{1}{T}\sum_i 
\left[-T^2 \frac{\partial^2}{\partial x_i^2} + \frac{1}{4} E_{,i}^2 - \frac{T}{2} E_{,ii}
\right]
 + \sum_{ij} E_{,ij} a^\dag_j a_i = (H^h)^\dag.
\label{hhhsusy}
\end{eqnarray}
$H^h_{FP}$ has now the standard form of a Schr\"odinger operator (acting
in imaginary time) with $T$ 
playing the role of $\hbar$. On the other hand $H^h$ is the standard 
Hamiltonian of  Supersymmetric Quantum Mechanics \cite{Cooper95}.

As the original operators $H_{FP}$ and $H$ are not hermitian we will have two
different eigenvalue equations, one for the right eigenstates 
($\ket{\bol{\psi^R}}$) and one for the left eigenstate ($\bra{\bol{\psi^L}}$)
\begin{equation}
H \boldsymbol{ | \psi^R \rangle} = \lambda 
\boldsymbol{ | \psi^R \rangle}   \;\;\;\; ; \;\;\;\;
\boldsymbol{  \langle  \psi^L|} H  =  \lambda 
\boldsymbol{  \langle  \psi^L|},
\end{equation}
while in the Hermitian basis there will be only one equation
\beq
H^h\ket{\bol{\psi^h}}=\lambda \ket{\bol{\psi^h}}.
\eeq
The three states are related by
\begin{equation}
 \boldsymbol{ | \psi^R \rangle} = e^{-\beta E/2} 
 \boldsymbol{|\psi^h \rangle   } 
\;\;\; ; \;\;\;
\boldsymbol{ | \psi^L \rangle} = e^{\beta E/2} 
 \boldsymbol{|\psi^h \rangle}.
\label{0sta}
\end{equation}
It is clear that $H$ and $H^h$ have the same spectrum.
Furthermore, the relation:
\beq
H^\dag = \sum_i \frac{\partial}{\partial x_i} \left[T \frac{\partial}{\partial x_i} - E_{,i} \right]
+\sum_{ij} E_{,ij} a_i a^\dag_j,
\label{gogo}
\eeq
implies that a left $k$-fermion eigenstate of $H$ is an $N-k$ right eigenstate
of the problem with the inverted potential $-E$.

 From (\ref{hsusy}) and (\ref{hhsusy}), we see that $H$ and $H^h$ have 
non-negative eigenvalues. 
By construction, there is at least one eigenstate $\ket{\bol{\psi^{0h}}}$ 
corresponding to the eigenvalue $\lambda=0$ (the smallest possible):
\begin{eqnarray}
\ket{\bol{\psi^{0h}}} &\propto& e^{-\beta E/2}\otimes \vidk, \nn
\ket{\bol{\psi^{0R}}} &\propto& e^{-\beta E} \,\,\, \,\otimes\vidk, \nn
\ket{\bol{\psi^{0L}}} &\propto& \mathrm{constant} \,\,\vidk. 
\label{rightleft} 
\end{eqnarray}
In order for $\ket{\bol{\psi^{0h}}}$ to be normalizable  we need 
the convergence of the integral (\ref{integre}).

The left and right eigenstates (\ref{rightleft}) have zero fermions and, 
thus, they belong also to the spectrum 
of $H_{FP}$. Clearly, both $Q$ and  ${\bar Q}$ annihilate 
$\ket{\bol{\psi^{0R}}}$. 
It is  easy to show (see Appendix \ref{qrel}) that this is necessary for any zero 
eigenvector,
and indeed (\ref{rightleft}) are the only ones with this property {\em if the space  has no holes}. 

In general applying $Q$ to any eigenstate   
$\boldsymbol{ | \psi^R \rangle}$ we get either a degenerate
eigenstate with one less fermion
 or zero. Similarly,  applying ${\bar Q}$
 we get either a degenerate eigenstate with one more fermion
 or zero.
 Each non-zero energy eigenstate $\ket{\bol{\psi^R}}$, 
annihilated by $Q$, can be written as $\ket{\bol{\psi^R}}=Q \ket{\bol{\chi^R}}$,
and the same holds for $\bar Q$. Indeed, from the eigenvalue equation
\beq
H\ket{\bol{\psi^R}}= \frac{1}{T}({\bar Q}Q+Q {\bar Q})\ket{\bol{\psi^R}} =
\frac{1}{T}Q{\bar Q}\ket{\bol{\psi^R}}=\lambda \ket{\bol{\psi^R}},
\eeq
one can infer that  
\beq
\ket{\bol{\chi^R}}=\frac{1}{T\lambda}{\bar Q}\ket{\bol{\psi^R}},
\label{qmat}
\eeq
satisfies $  Q \ket{\bol{\chi^R}} =\ket{\bol{\psi^R}}$.  

In conclusion, each non-zero energy eigenstate with $k$ fermions, 
will have one and only one  supersymmetric partner with either $k-1$ or $k+1$  fermions.
In a space with no holes the only eigenstate with zero energy
has zero fermions and  is  the Gibbs measure (See Appendix \ref{qrel}).
 The spectrum is organized
as in  Fig. \ref{spectrumn}.

\begin{figure}
\centering \includegraphics[totalheight=6cm]{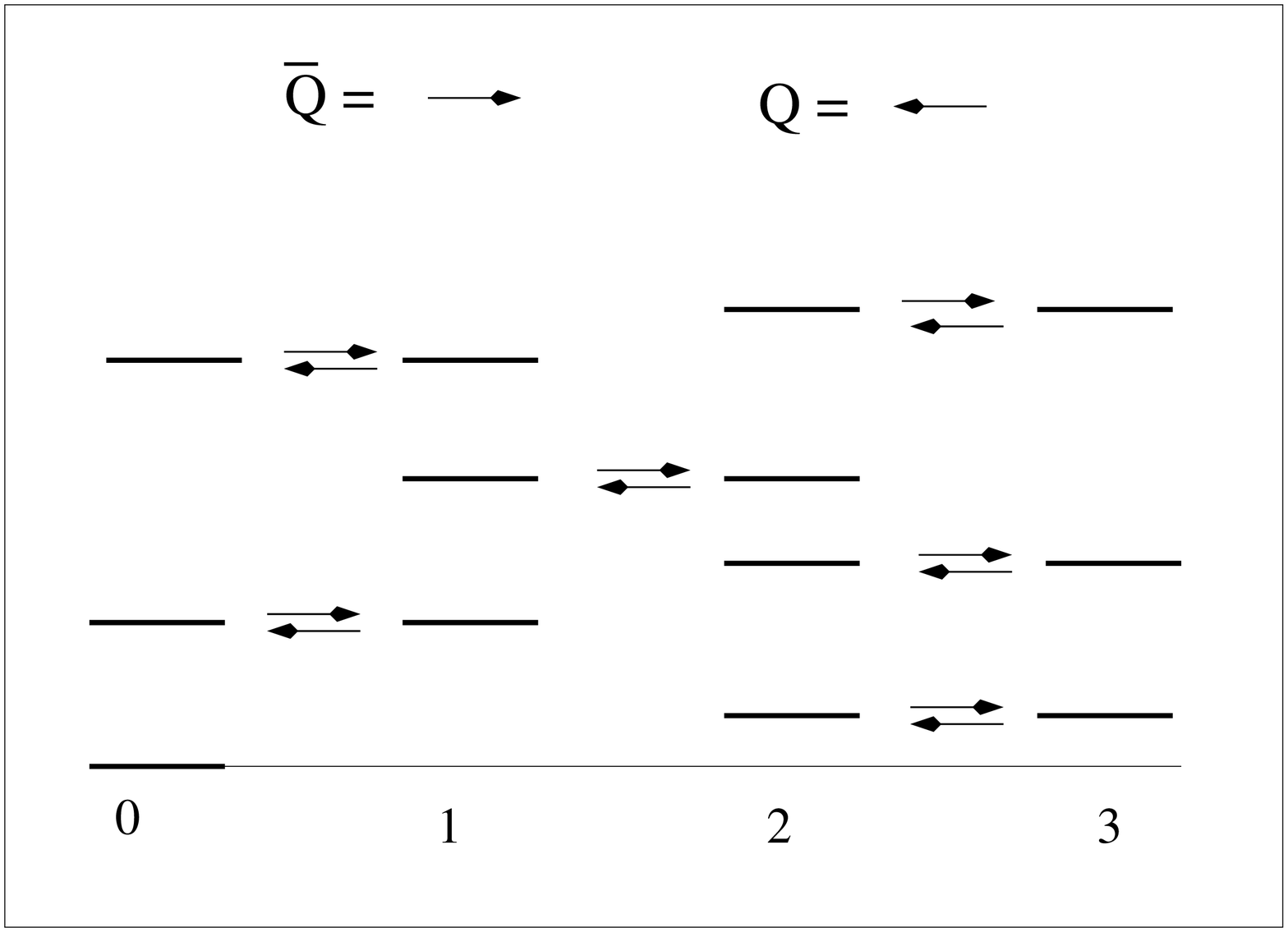}
\caption{
The pairing of the energy levels for a generic spectrum. 
Each eigenstate of positive energy has a supersymmetric partner. The number 
of fermions is written below each corresponding column. The only unpaired
state is the zero-energy one.
}
\label{spectrumn}
\end{figure}


\section{Morse theory}
\label{sec:morse}
In the low temperature limit, the organization of the spectrum of $H^h$ allows
 to derive relations concerning the critical points (saddles) of the energy 
surface, defined as those points for which
\beq
|{\boldsymbol \nabla}E|^2=\sum_{i=1}^N E_{,i}^2=0.
\eeq

Let us study the {\em semiclassical} low-$T$ spectrum.
At the lowest order in $T$, the potential  appearing in the Schr\"odinger 
operator (\ref{hhsusy})
\beq
W=\frac{1}{T}\sum_i \frac{1}{4} E_{,i}^2,
\eeq
is very large except at the critical points of $E$;
 so the  eigenstates  of $H^h$ with low-lying eigenvalues
are concentrated around the 
those critical points.

Let us assume that the critical points are isolated and the Hessian $E_{,ij}$
 has non zero 
eigenvalues. Then, as usual, the semiclassical development
starts with a harmonic approximation around each minimum  of $W$.
Consider one of these minima, where the Hessian has eigenvalues  $A_1,\dots,A_N$. 
We can develop $E$ in the local coordinates,
and going to the basis in which the  Hessian $E_{,ij}$ is diagonal, we have
that, locally:
\beq
E(\bol{x'}) \sim E_0+\frac{A_i}{2}x'^2_i.
\eeq
We can develop $H$ at the first order in $T$ as
\begin{equation}
\frac{T}{2} H' = \sum_i\left\{ - \frac{T^2}{2} \frac{\partial^2 }{\partial x^2_i} + \frac{1}{2} 
\left(\frac{A_i}{2}\right)^2 x'^2_i   - \frac{T}{2}A_i
+T A_i a'^{\dag}_i a'_i \right\}.
\label{quadr}
\end{equation}
We recognize  the Hamiltonian of  $N$ independent 
oscillators plus $N$ independent fermion terms.
Along each direction $i$ on the right hand side, we have a harmonic oscillator
with {\em positive} frequency $|A_i|$, plus terms which give  $-A_i/2$ if there
is no fermion and $+A_i/2$ if there is a fermion along the direction $i$.
Hence, each fermion term will exactly cancel the zero-point energy of each oscillator, provided
we have
  zero fermions if  $A_i>0$ and  one fermion if $A_i<0$.
All in all, we see that we get zero to this order if and only if we have exactly as many
fermions as unstable directions in the particular critical
 point \cite{endnote3}.

The next higher eigenvalues $\lambda$ are  given by, to leading order:
\begin{equation}
\lambda \sim  \frac{1}{2} \sum_i \left[(2N_i+1)|A_i|-A_i + 2A_i n_i)\right],
\label{quadr2}
\end{equation}
with $N_i=0,1,2,\dots$ and $n_i$ is the number of fermions ($n_i=0,1$) for 
each  direction. The spatial part of the eigenstates are Gaussians times 
polynomials, thus having widths of order $\sqrt{\frac{T}{A_i}}$ in the 
$i^{th}$ 
direction so the approximation is consistent at low $T$.

Let us call saddle of index $p$ a critical point  whose Hessian has $p$ 
negative eigenvalues,  and  $M_p$ the number of these. For example 
the minima are saddles of order $0$ while the saddles of order $N$ are 
maxima. From the above considerations  it follows that around
each saddle of index $p$ there there is one and only one state with  zero 
energy as $T \rightarrow 0$,  and this state has $p$ fermions. 
This means that the Hamiltonian (\ref{hhsusy}) has, to this order,
 $M_p$ $p$-fermion eigenstates with a zero energy.
 As a consequence, there is a gap in the eigenvalues of each fermion sector, and  
the spectrum looks like Fig. \ref{spectrum} \cite{endnote4}.

Recalling that any non-zero energy eigenstate with $p$ fermions has a
degenerate partner with either $p-1$ or $p+1$ fermions (cfr. Section I),   
one can read from Fig. \ref{spectrum} the relations:
\bea
M_0&=&1+K_1,\nn
M_1&=&K_1+K_2,\nn
&\vdots &\nn
M_N&=&K_N.
\label{morser}
\eea
The positivity of  the $K$'s ( $K_i\geq0 \quad \forall i$) 
are the strong Morse inequalities.

\begin{figure}
\centering \includegraphics[totalheight=6cm]{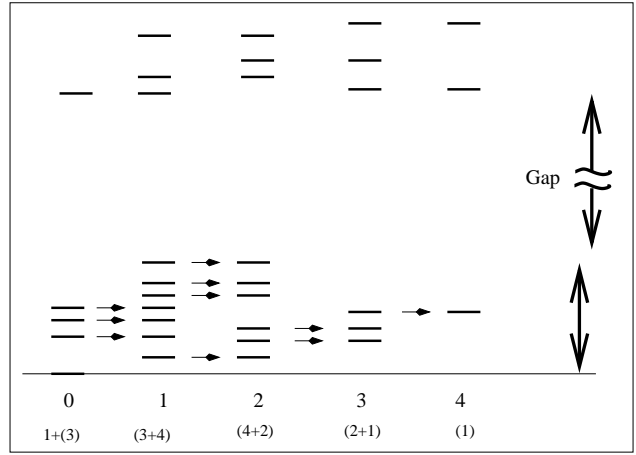}
\caption{
Morse Theory. The arrows indicate the action 
of ${\bar{Q}}$. The gap in the spectrum means that the ratio of the lowest
eigenvalue above the gap to the highest below the gap becomes
infinite at small $T$.
The numbers between brackets indicate the number of states below the
gap of each fermion number decomposed as in (\ref{morser}); 
the Morse inequalities are evident from the picture. 
}
\label{spectrum}
\end{figure}

We have used the condition that $E$ is defined
 in a space without holes. As we have seen (see
Appendix \ref{qrel}) this implies that there is only one eigenstate of zero
energy and it has zero fermions. If the space  has a more complicated topology, 
 there will be several zero-energy eigenstates not paired
by the supersymmetric charges, and the Morse inequalities become 
slightly more complicated (see Appendix \ref{exotic}).

\section{States and transition Currents}

\label{sec:states}

\subsection{A Simple Case}
\label{sec:simple}

In the previous section we have given the form of the spectrum, at least to leading order, 
and the corresponding wavefunctions  of every fermion number -- the latter in the Hermitian basis.
It may seem that going to the original basis is trivial since it 
is simply a matter of multiplying
those approximate wavefunctions by $e^{\beta E/2}$. As mentioned above, this is rather tricky,
since the factor $e^{\beta E/2}$ will resurrect large deviations which we have neglected.

Let us first study the simple case of a double well at low temperatures.
 We consider a probability distribution
 $P$ evolving under the action of the Fokker-Plank Hamiltonian (\ref{fokker})
corresponding to an energy as in Fig. \ref{exemple}. 
This distribution  can be decomposed on the eigenstates of the Fokker-Planck Hamiltonian (ie. the 
zero-fermion eigenstates of the Hamiltonian (\ref{hsusy}))
\beq
P(x,t)=\sum_{\alpha=0} c_\alpha \psi^{\alpha R} e^{-\lambda_\alpha t},
\eeq
where $\psi^{0R}=e^{-\beta E}$ while $\lambda_0=0$.

\begin{figure}
\centering \includegraphics[totalheight=6cm]{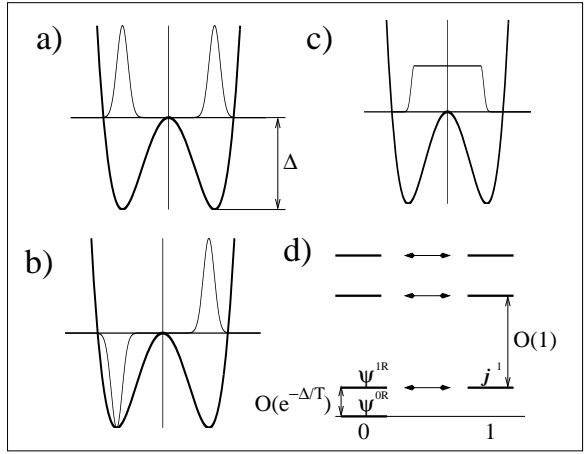}
\caption{
The potential and different eigenstates along the reaction coordinate;
(a) is the equilibrium density ($\psi^{0R}$), (b) the first eigenstate - the most stable 
($\psi^{1R}$), (c) is the current density in the first eigenstate $j^1$ and
(d) the spectrum with the gap and the two fermionic sectors. }
\label{exemple}
\end{figure}

In the low temperature limit  there are  
  two metastable states each concentrated around
one of the minima. Barrier penetration leads to the Gibbs measure,
the symmetric combination of those states.
 In fact the spectrum of the Fokker-Planck Hamiltonian will contain one zero eigenvalue 
$\lambda_0=0$ (the Gibbs measure), one small $\lambda_1 \sim O(e^{-\Delta/T})$
eigenvalue  and the rest of them 
much larger ($O(1)$). The two pure states, localized on the right and on the left 
 are $\propto \; \psi^{0R}(x) \pm \psi^{1R}(x)$, respectively.
 If we are interested in the dynamics of the 
passage  between the two wells we have to consider times such that the fast 
relaxation within each well has already taken place. 
At such times, larger than 
$t_1\sim \frac{1}{\lambda_2}log(\frac{c_2}{c_1})$, 
 we are left only with a distribution 
\beq
P(x,t \gg t_1) \simeq c_0 \psi^{0R}+c_1\psi^{1R}e^{-\lambda_1 t},
\label{density}
\eeq
i.e. a combination of states localized to the right and to the left,
 dependent upon the initial condition
and time.

The current at any time is given by:
\beq
j(x,t)=\left(T\frac{\partial}{\partial x}+
\frac{\partial E}{\partial x}\right)P(x,t)= \sum_{\alpha=1} c_\alpha e^{-\lambda_\alpha t}
\left(T\frac{\partial}{\partial x}+\frac{\partial E}{\partial x}\right)\psi^{\alpha R},
\label{curr}
\eeq
and its divergence reads: 
\beq
\frac{\partial j}{\partial x} = 
\frac{\partial }{\partial x}\left(T\frac{\partial}{\partial x}+
\frac{\partial E}{\partial x}\right)P(x,t) =
 - \sum_{\alpha=1}\lambda_\alpha c_\alpha e^{-\lambda_\alpha t}\psi^{\alpha R}.
\label{divv}
\eeq
We can split the contribution of each term as:
\beq
I^\alpha \equiv \lambda_\alpha c_\alpha e^{-\lambda_\alpha t}\psi^{\alpha R}.
\eeq
All the contributions except $I^1$ eventually vanish,  for example:
\beq
\frac{I^2}{I^1}=\frac{c_2\lambda_2e^{-\lambda_2 t}}{c_1\lambda_1e^{-\lambda_1 t}} \ll 1,
\eeq
 at times such that
 $t  \gg t_2 \sim \frac{1}{\lambda_2}log(\frac{c_2\lambda_2}{c_1\lambda_1})$
\cite{endnote5}.

Equation (\ref{divv}) now implies that the late-time regime current $J^1(x)$ is:
\beq
J^1(x) = c_1  \lambda_1 e^{-\lambda_1 t} \int_x^{\infty} dx \, \psi^{1R}(x)=c_1 j^1(x)e^{-\lambda_1 t},
\eeq
and this in encoded in the one-fermion right eigenstate `below the gap' as:
\beq
\ket{\bol{\xi^{1R}}}={\bar Q}\ket{\bol{\psi^{1R}}}=  -\mathrm{i} \int dx \, j^1(x)\;  a^\dag \ket{x} \otimes \vidk.
\eeq
In Fig. \ref{exemple}  we summarize the situation.
Note that  although the state $\ket{\bol{\xi^{1h}}}$
is sitting on the saddle (as we have seen in the previous section), its
form in the original basis $ \ket{\bol{\xi^{1R}}} = e^{-\beta E/2}\ket{\bol{\xi^{1h}}} $,
which encodes  the current, is essentially a constant between the two wells.

\subsection{States}
\label{sec:states1}

We have seen in the previous example how in the low temperature limit
one can unambiguously define metastable states using the eigenstates of the 
Fokker-Planck operator.
 When the energy function is rough, with several non-equivalent minima, or when the
origin of metastability is not the low temperature,
the construction is less obvious.
Suppose the Fokker-Planck spectrum has $K$ eigenstates with
 low eigenvalues $0,\lambda_1,\dots,\lambda_{K-1}$, separated by 
a gap from all the higher ones. 
One can show \cite{endnote1} that, to the 
extent that the gap is large 
$(\lambda_K-\lambda_{K-1}\gg \lambda_{K-1})$, one can
construct {\em exactly} $K$ distributions $P_0(\bol{x})$, ..., $P_{K-1}(\bol{x})$
 by linear combinations of the 
right eigenstates of the Fokker-Plank Hamiltonian  $\psi^{\alpha  R}(\bol{x})$ with 
$0\leq \alpha < K$:
\begin{equation}
P_\alpha(\bol{x}) = \sum_{\gamma=0}^{K-1} T_{\alpha  \gamma } \psi^{\gamma  R}(\bol{x}),
\label{GS1}
\end{equation}
such that  the $P_\alpha(\bol{x})$ are either positive or negligible, and mutually
disjoint (the  product of any two is everywhere negligible), 
 these are the 
{\em states}.
We shall take the $P_\alpha(\bol{x})$ normalized 
$\int d^N x\; P_\alpha (\bol{x})=1 \;\; \forall \; \alpha$. 
Every combination of the right eigenstates `below the gap'
can be expressed as a linear combination of 
the states. In particular, 
the Gibbs measure is:
\begin{equation}
\psi^{0R}(\bol{x}) = \sum_{\alpha =0}^{K-1} T_{0\alpha}^{-1} P_\alpha(\bol{x}).
 \end{equation}
A useful formula is obtained integrating (\ref{GS1}) with respect to
$\bol{x}$, and noticing that $\brak{\bol{\psi^{0L}}}{\bol{\psi^{\alpha  R}}}=0 \;\; \forall \;
\alpha >0$:
\begin{equation}
T_{\alpha  0}=1 \;\;\; \forall \; \alpha. 
\label{T01}
\end{equation}

The left eigenstates `below the gap' are also interesting. By linear combinations of the 
left eigenstates  $\psi^{\alpha  L}(\bol{x})$ with $0\leq \alpha < K$:
\begin{equation}
A_{\alpha}(\bol{x}) = \sum_0^{K-1} T_{\alpha \gamma } \psi^{\gamma  L}(\bol{x}), 
\label{GS2}
\end{equation}
one obtains functions $A_\alpha(\bol{x})$, ..., $A_{K-1}(\bol{x})$ such that each 
$A_\alpha(\bol{x})$ is essentially constant where $P_\alpha$ is non-negligible, and is negligible 
elsewhere.  To summarize,  right eigenstates below the gap are locally Gibbsean, while
 the corresponding left ones are essentially constant within a `state'. 
The case of very low temperatures is the the simplest one to visualize:
in this case the $P_\alpha$ are Gaussians sitting each one at the bottom of a local minimum,
and the $A_l(\bol{x})$ are constant within the corresponding (zero temperature) basin of attraction
of each minimum, and zero elsewhere.

The low-lying eigenvalues can also be interpreted as exit times.  
 Under the assumption of well separated eigenvalues,
their inverses  
give the exit times corresponding to the metastable states. In fact, the 
low-lying eigenvalues, together with their corresponding eigenstates will 
completely define the long-time dynamics of the system. Indeed, 
 one can define, from a probability density $P(\bol{x})$, 
the site-populations as
\beq
c_\gamma =\int d^N \!\! x\; A_\gamma (\bol{x})P(\bol{x}),
\eeq
and obtain a master equation for $c_\gamma (t)$ as
\beq
\frac{d c_\gamma}{d t}=\sum_{\nu=0}^{K-1}w_{\gamma \nu}c_\nu \quad w_{\gamma \nu}=
\braok{\bol{A_\gamma}}{H_{FP}}{\bol{P_\nu}}.
\label{master}
\eeq
This master equation is accurate at times for which 
$\lambda_{K}e^{-\lambda_{K}t} \ll \lambda_{K-1}e^{-\lambda_{K-1}t} $.

Let us conclude by mentioning that there are in cases in which there are more than two separated timescales
one can make this construction at more than one level -- thus obtaining different sets of
states relevant for the different timescales.

\subsection{Transition currents from the 1-fermion eigenstates}
\label{sec:tran}

Let us assume that the Fokker-Planck spectrum has exactly $K$ eigenstates `below the gap',
implying that there are $K$ metastable states.
We shall show that, whatever the origin of the gap,
 exactly as in section \ref{sec:simple} the $K-1$ one-fermion partners
of these eigenstates are the reaction current
distributions at long times (for which 
$\lambda_{K}e^{-\lambda_{K}t} \gg \lambda_{K-1}e^{-\lambda_{K-1}t} $).
This may seem rather surprising: the construction of states in the previous
section is not a priori
good in the regions where the probability is negligible. This is not important
 at the level of the probabilities, as those regions carry vanishingly small
weight. However, it seems to pose a problem for the current, which
  is important on the  barrier -- in which region the  
 probability corresponding to a state  small -- and drops to zero 
 within a state,
 precisely where the approximation described in the previous section is
reliable.
We shall argue now that, in spite of this apparent limitations,
 as we have already seen in the section 
\ref{sec:simple}, the $1$-fermion partners give exactly the long-time currents
between states.

Let us make the following gedanken-experiment. At time zero we 
prepare a probability density such that it will fall into one 
state, say $\alpha_o$, $ 0 \leq {\alpha_o} \leq K-1$ . 
The initial  probability density can then be written as
\beq
P(\bol{x},0)=\sum_\alpha c_\alpha  \psi^{\alpha R}=c P_{\alpha_o}   (\bol{x})+\sum_{\alpha\geq
K} c_\alpha \phi^{\alpha  R}.
\eeq

We shall  study $P(\bol{x},t)$ at a time $t$ in the middle of the  gap, 
that is $\lambda_{K}t \gg 1 \gg \lambda_{K-1} t$, a time sufficiently  large  so that 
 all the fast components become very small but not yet  large enough as to  
populate  other states. The density is then
\beq
P(\bol{x},t)= c P_{\alpha_o}(\bol{x})+O(e^{-\lambda_{K}t})+O(1-e^{-\lambda_{K-1}t}).
\label{decref}
\eeq

Let us study now the current 
\beq
J_k=\left(T\frac{\partial}{\partial x_k}+
\frac{\partial E}{\partial x_k}\right)P=\sum_{\alpha=1}
 c_\alpha e^{-\lambda_\alpha t}\left(T\frac{\partial}{\partial x_k}+
\frac{\partial E}{\partial x_k}\right)\psi^{\alpha R} = \sum_{\alpha=1} j_k^{\alpha} ,
\eeq
where we have discriminated the contributions to the current  of each eigenstate
\bea
j^{\alpha}_k \equiv  c_\alpha e^{-\lambda_\alpha t}\left(T\frac{\partial}{\partial x_k}+
\frac{\partial E}{\partial x_k}\right)\psi^{\alpha R},
\eea
(using the notations from the previous section and (\ref{decref}) one can see that 
$c_\alpha=c T_{\alpha_o \alpha}$).
The  $\bol{j^{\alpha}}(\bol{x})$ are not normalized, so it is difficult to
compare them. In order to do so, we compute the divergence of the corresponding terms:
\beq
\mbox{div}\; \bol{j^{\alpha}}(\bol{x}) =
  c_\alpha e^{-\lambda_\alpha t} \lambda_\alpha \psi^{\alpha  R}(\bol{x}).
\eeq
The relative contribution to  the current of two terms is of the order:
 \beq
\frac{\int_{{\cal{V}}_\alpha} d^N\!\!x\;
\mbox{div}\; \bol{j^{\alpha}}(\bol{x})}{\int_{{\cal{V}}_\beta}d^N\!\!x\;
\mbox{div}\; \bol{j^{\beta}}(\bol{x})} \propto \frac{c_\alpha \lambda_\alpha}{c_\beta \lambda_\beta}
 e^{-t(\lambda_\alpha-\lambda_\beta)}.
\eeq
where ${\cal{V}}_\alpha$ and ${\cal{V}}_\alpha$ are the regions in which 
$\psi^{\alpha R}>0$, $\psi^{\beta R}>0$ respectively \cite{endnote6}.

Hence, for large enough times $t(\lambda_\alpha-\lambda_{K-1}) \gg 
\ln(\frac{c_\alpha \lambda_\alpha}{c_{K-1} \lambda_{K-1}})$, all states $\alpha$ above 
the gap do not contribute to the current.

In conclusion, we have shown that the escape current  of any metastable state is
a linear combination of the currents associated to  states below the gap.
This in turn means that, within this late-time regime, the current is encoded as
a linear combination of some of the one-fermion eigenstates `below the gap', those that have
zero-fermion partners
\beq
\int d^N\!\!x \; J_k(\bol{x})  a^\dag_k | \bol{x} \rangle \otimes | - \rangle  =
\mathrm{i}\sum_{\alpha=1}^{K-1}c_\alpha e^{-\lambda_\alpha t} \ket{\bol{\xi^{\alpha R}}}=
\mathrm{i}\sum_{\alpha=1}^{K-1}c_\alpha e^{-\lambda_\alpha t} \bar Q\ket{\bol{\psi^{\alpha R}}}.
\eeq
In the next section we interpret those having a two-fermion partner.

\subsubsection{Transition times}
\label{sec:trant}
Suppose one has the current $\bol{J}$ escaping a metastable state $P(\bol{x})$:
\beq
J_i(\bol{x}) \propto \left(T\frac{\partial}{\partial x_i}+E_{,i}\right)P(\bol{x}) \;\;\; ; \;\;\; 
P(\bol{x}) = \sum_{\alpha=1}^{K-1} c_\alpha \psi^{\alpha R}(\bol{x}).
\eeq
We wish to give an expression for the transition time in terms of the unnormalized
current $\bol{J}$. For this, we first compute:
\begin{eqnarray}
\int d^N\!\!x \; e^{\beta E} \; {\bol{J}}^2 &=&
 \left\{\left(T\frac{\partial}{\partial x_i}+E_{,i}\right)P \right\}
 \; e^{\beta E} \;
 \left\{ \left(T\frac{\partial}{\partial x_i}+E_{,i}\right)P \right\}
\nonumber \\
&=&
\int d^N\!\!x \; \left\{
\left(T\frac{\partial}{\partial x_i}+E_{,i}\right)P \right\}
 \; \frac{\partial}{\partial x_i}\left(e^{\beta E}P\right) \nonumber \\
&=&
\int d^N\!\!x \; P \; e^{\beta E} \; H_{FP}P = \sum_{\alpha=1}^{K-1}\lambda_\alpha c^2_\alpha,
\end{eqnarray}
and similarly:
\beq
\int d^N\!\!x\;
e^{\beta E}\;(\mbox{div}\; \bol{J})^2 = \int d^N\!\!x \,(H_{FP}P)\;e^{\beta E}\;(H_{FP}P)
=\sum_{\alpha=1}^{K-1}\lambda_\alpha^2 c_\alpha^2,
\label{summ}
\eeq
because the sums are dominated by the largest eigenvalues $\lambda_{max}$ 
 within the sum (\ref{summ})
that contribute to the state $P(\bol{x})$, we have that the smallest escape time is:
\beq
t_{activ}=\lambda_{max}^{-1} = \frac{\int d^N\!\!x \; e^{\beta E} \; {\bol{J}}^2}
{\int d^N\!\!x \;
e^{\beta E}\;(\mbox{div}\; \bol{J})^2}.
\label{timescale}
\eeq
Note that the normalization of the current is irrelevant.
{\em This formula is valid on the assumption of separation of timescales,
irrespective of its cause.}
The Kramers expression for the low-temperature case can be easily
read of this formula, since the numerator is dominated by the exponential
of the barrier height and the numerator by the exponential of the energy 
 of the starting well.
If the current is divergence-less (a loop, as  we shall encounter later), 
the timescale is infinite.

An immediate conclusion one draws from (\ref{timescale}) is that if one knows
the current with an error $\delta  {\bol{J(x)}}$, it is in the regions with large energy
(the saddles) and with large divergence (the starting region) where this error
translates into a larger error in the timescale.

\subsection{Loops: blind saddles and subdominant paths}
\label{sec:loops}
The one-fermion sector contains in general two kinds of eigenstates states
`below the gap'. These are those  given by ${\bar{Q}}$ acting 
on a zero-fermion state, and those given by ${{Q}}$ acting 
on a two-fermion state. The former give us the dominant  reaction currents, as we have seen already.
We now show that the latter give us current loops, and in particular the 
the alternative (subdominant) routes between states. 

The states we are now considering are constructed as follows: given a two-fermion
eigenstate
$\boldsymbol{ |\rho^R  \rangle}= \sum_{ij} a^\dag_ia^\dag_j| \rho^R_{ij} \rangle $, 
we obtain a one-fermion eigenstate as:
\begin{equation} 
{ Q}\boldsymbol{ | \rho^R \rangle} \equiv 
\boldsymbol{ |\chi^R  \rangle}= a^\dag_i \ket{\chi^R_i} \otimes  \vidk; \quad
\chi^R_i(\bol{x}) = -\mathrm{i} T
 \sum_j  \left(\frac{\partial \rho^R_{ij}}{\partial x_j}-\frac{\partial \rho^R_{ji}}{\partial x_j}\right),
\end{equation}
unless $\ket{\chi^R_i} =0$.
From ${Q}\boldsymbol{ |\chi^R  \rangle}=0$ we immediately
conclude that field of the right eigenstate is divergenceless:
\begin{equation}
\sum_i \frac{\partial \chi^R_i(\bol{x})}{\partial x_i} = 0.
\end{equation}
so that if $ \ket{\bol{\chi^R}}$ encodes a single current line, it must be a closed loop. 

\begin{figure}
\centering \includegraphics[angle=-90,totalheight=6cm]{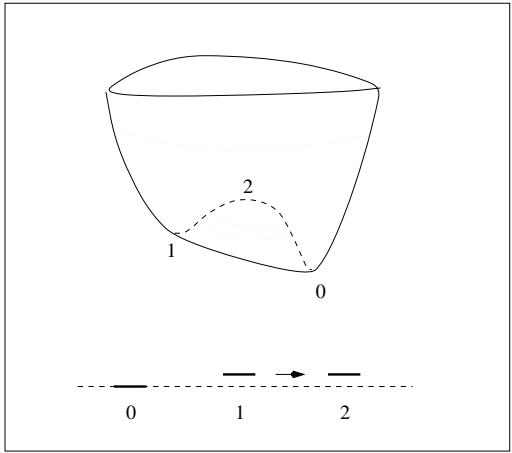}
\caption{A landscape with a minimum, a maximum and a blind
  saddle. Below: the low eigenvalue spectrum for zero, one and two
  fermions. The dotted line is the zero level, 
other eigenvalues are exponentially small in $1/T$.
Next higher eigenvalues start at $O(1)$ (not shown). 
}
\label{mexican}
\end{figure}

\subsubsection{Simple examples}
\label{sec:simplee}
A low temperature example will make things clearer.
Consider the tilted Mexican hat in two dimensions Fig. (\ref{mexican}): 
it has a minimum, a maximum, and
a `blind' saddle, one that does not lead anywhere.
The two-fermion lowest eigenstate is of the form
\begin{equation}
 |\boldsymbol{ \rho^R }\rangle =  a^\dag_x a^\dag_y \ket{\rho^R}\otimes \vidk,
\end{equation}
where $\rho^R(x,y)$ satisfies:
\begin{equation}
- \left(T\frac{\partial}{\partial x}+
\frac{\partial E}{\partial x}\right)\frac{\partial}{\partial x}
\rho^R(x,y)- \left(T\frac{\partial}{\partial y}+
\frac{\partial E}{\partial y}\right)\frac{\partial}{\partial y}
\rho^R(x,y)=\lambda \rho^R(x,y),
\label{frrr}
\end{equation}
which is easily obtained permuting (fermion) particles and holes
in the  Hamiltonian (\ref{hsusy}).
The lowest-lying one-fermion eigenstate is obtained
by noticing that the eigenvalue equation (\ref{frrr}) corresponds to
the equation satisfied by the {\em left} eigenstate of a
Fokker-Planck
equation in a the reversed potential $-E(x,y)$  (cfr. Eq. (\ref{gogo})).
From the discussion in \ref{sec:states}, we 
  conclude
that $\rho^R$ (the only $A(x,y)$ for the reversed problem) is essentially constant
within the region spanned by all gradient lines descending from the
local maximum (the unstable manifold of the maximum,
or the stable manifold of the minimum  of $-E$)
 -- and drops sharply to zero at the border of this region.
Acting with $Q$ on $\rho^R$, we obtain the current:
\begin{equation}
\left( \chi^R_x(x,y),\chi^R_y(x,y) \right) \sim 
\left(\frac{\partial \rho^R}{\partial y},- \frac{\partial \rho^R}{\partial
  x} \right), 
\end{equation}
which is then non-negligible on the gradient  paths joining the minimum with the saddle,
because this is where  $\rho^R$ has a non-negligible gradient.
The direction is turnaround, and  
clearly the flow so obtained is divergence-free.

Let us now see the general relation between 
passages and loops with another slightly more complicated low temperature example.
Consider a situation as in Fig. \ref{morsem}. There are four minima, multiply
connected by seven paths going through as many saddles.
At low temperature, only three of them (shown in thicker lines) have a much shorter
passage time, and hence dominate the reactions.
The other four can be obtained from combination of these and the four independent loops
--- for example one can take each loop going around each of the four  maxima.
The eigenstate structure `below the gap' reflects this: there are four zero-fermion
(right) eigenstates corresponding to four minima. One of them is the Gibbs measure,
the other three have one-fermion partners yielding the three dominant passages.
The remaining four one-fermion (right) eigenstates correspond to the loops, and they have
two-fermion partners corresponding to the regions they encircle, including
each a  maximum. 

 \begin{figure}
\centering \includegraphics[totalheight=6cm]{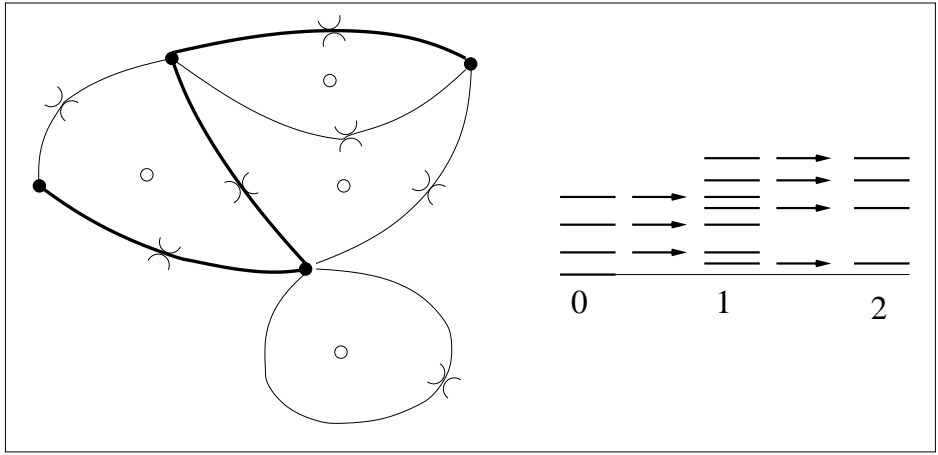}
\caption{
A sketch of an energy surface with four minima  (full circles), four 
maxima (open circles) and seven pathways passing through one saddles.
The thick paths have a low activation times. On the right the corresponding
spectrum of the Hamiltonian (\ref{hsusy}).}
\label{morsem}
\end{figure}

The simple tilted Mexican hat problem, and in general any  two-dimensional
situation,  allows us also to understand the different roles played by 
 partner eigenstates below the gap in the one and two fermion subspaces.
 As we have mentioned above, all the right two-fermion eigenstates which are
partners to the loops  can be obtained
(always in two dimensions) from the 
zero-fermion left eigenstate of the inverted potential.
This means that each corresponds to a constant in the region spanned by all trajectories 
descending from a saddle of index two (its unstable manifold),
 and this will be also true in more dimensions.

\subsubsection{Loops: physical meaning and derivations} 
\label{sec:loopsp}
The loops have 
also a  physical meaning which may be extended to apply to
the nonzero temperature situation.
 Consider a system in equilibrium to which we add
a force field $h \bol{f}(\bol{x})$, $h$ small,
that has only rotational in a restricted region ${\cal{D}}_\Gamma$ of phase space:
\begin{equation}
 \frac{\partial f_j}
{\partial x_k}-\frac{\partial f_k}{\partial x_j} = 0 \;\;\; \forall j,k \;\;
if \;\; \bol{x} \notin {\cal{D}}_\Gamma.
\end{equation}
The effect of such a field will be to create currents  which will persist even in the stationary state.
 In a system with metastability, 
these currents can be of two types: those generated essentially within a state, and 
those due to forced passages through barriers; the latter are  the loops.
We shall see that the currents within a state  are given by 
eigenstates above, and the loops by eigenstates below the gap of the one-fermion spectrum.

To make this clear let us go  back to the tilted Mexican hat (Fig. 
\ref{mexican}). Let us consider a force whose rotational is concentrated
in a restricted `vorticity' region ${\cal{D}}_\Gamma$ (the dark region in the figure). 
If the vorticity is concentrated close to the minimum (\ref{mexican}a) the  currents 
generated will be due to particles which in a rare excursion happen to 
fall upon ${\cal{D}}_\Gamma$, and then typically fall right back to the state. If we shift 
the ${\cal{D}}_\Gamma$ further away from the state, we get a behavior of the same kind until
we reach a point in which the vorticity is located higher than the saddle point, and
it becomes more probable for the current to go round the saddle (\ref{mexican}b) through gradient 
lines: this is the loop distribution and its essentially independent of the exact 
position of ${\cal{D}}_\Gamma$.
 It is given by the (only) one-fermion eigenstate below the gap.
As we shall see below, the condition that the vorticity generates a loop around a saddle
is that it pierces  the surface on which the two-fermion eigenstate `below the gap' is
non-zero: this is a general fact.
 
\begin{figure}
\centering \includegraphics[angle=0,totalheight=6cm]{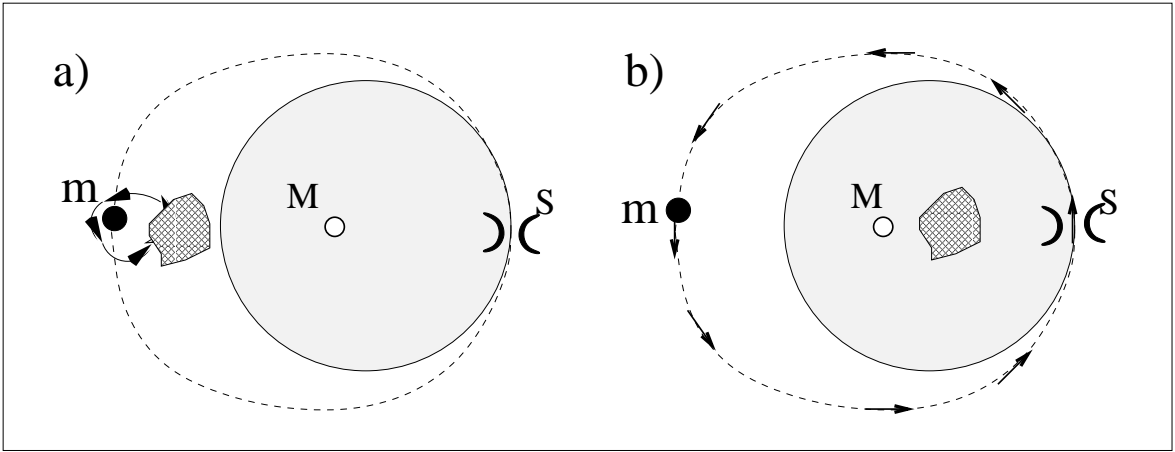}
\caption{ The tilted Mexican hat seen from above. The full lines are the 
level lines at the energy of the saddle (S):
the light gray region is above the  saddle level.
 The broken line corresponds to the gradient
line going from the minimum (m), through the saddle, and back to the minimum  encircling the 
maximum (M).  The dark gray
region corresponds to the domain ${\cal{D}}_{\Gamma}$ where the  
drift has a non-zero vorticity. Left: ${\cal{D}}_{\Gamma}$ below the saddle level - 
currents above the gap; Right:  ${\cal{D}}_{\Gamma}$ above the saddle level - 
currents below the gap. See text.}
\label{tiltgama}
\end{figure}

In  order to see this quantitatively, let us  study the perturbed Fokker-Planck equation:
\begin{eqnarray}
H^f_{FP} &=&-\sum_i\frac{\partial}{\partial x_i} \left(T\frac{\partial}{\partial x_i}+
 E_{,i}+h f_i\right),
\label{fokkerpp}
\end{eqnarray} 
with $h$ a small parameter and $f_i$ as above.
Proposing a stationary distribution of the form:
\beq
H^f_{FP}P_{st}=0 \;\;\; ; \;\;\; P_{st}=c_o e^{-\beta E}+h P^1,
\eeq
($c_o$ the normalization of the Gibbs measure)
the current in the stationary state is obtained as:
\begin{eqnarray}
J^{st}_i&=&\left(T\frac{\partial}{\partial x_i}+E_{,i}+h f_i\right)P_{st}=
h \left[c_o f_i e^{-\beta E} +\left(T\frac{\partial}{\partial x_i}+E_{,i}\right)P^1\right]
\nonumber \\ 
&=&h e^{-\beta E} \left[c_o f_i + T \frac{\partial}{\partial x_i} \left(e^{\beta E} P^1\right) \right],
\label{curre}
\end{eqnarray}
and is zero only if the square bracket vanishes, i.e. if $\bol{f}$ derives
from a potential.
The current is obviously divergence-free, and it is easy to see that 
the corresponding one-fermion state $\ket{\bol{\xi^R}}= 
\int d^N\!\!x \; \sum_iJ^{st}_i(\bol{x}) a^\dag_i \ket{\bol{x}}\otimes \vidk $ satisfies:
\begin{eqnarray}
Q \ket{\bol{\xi^R}} &=& 0, \label{gggt}
\\
{\bar{Q}} \ket{\bol{\xi^R}} &=& c_o h T e^{-\beta E} \ket{\bol{\Gamma}},
\label{annihilators}
\eea
where
\beq
 \ket{\bol{\Gamma}} =-\mathrm{i} \left(\frac{\partial f_j}
{\partial x_k}-\frac{\partial f_k}{\partial x_j} \right) a^\dag_ja^\dag_k\vidk.
\eeq

Equation (\ref{gggt}) implies that $\ket{\bol{\xi^R}}$ can be developed in terms
of one-fermion eigenstates with two-fermion partners.
Multiplying the Eq. (\ref{annihilators})  by $Q$, we obtain:
\beq
\ket{\bol{\xi^R}} = c_o h H^{'-1} Q e^{-\beta E} \ket{\bol{\Gamma}} =
 c_o h \; H^{'-1}e^{-\beta E} {\bar{Q}}^\dag  \ket{\bol{\Gamma}},
\label{jjj}
\eeq
where $H'$ is  $H$ restricted to the one-fermion subspace.
Developing (\ref{jjj}) in a basis, we find:
\beq
\ket{\bol{\xi^R}} =  c_o h \sum_{\alpha} \frac{1}{\lambda_\alpha}
 \ket{\bol{\xi^{\alpha R} }}\braok{\Gamma}{\bar Q}{\bol{\xi^{\alpha R}}}.
\label{kkk}
\eeq
Defining as $H_{+}$ and $H_-$ the projections of $H'$ above and below the gap,
respectively, we introduce the components of the current distribution:
\begin{eqnarray}
\ket{\bol{\xi^R_{state}}} &=&c_o h H^{-1}_+ e^{-\beta E} {\bar{Q}}^\dag  \ket{\bol{\Gamma}}, \nn
\ket{\bol{\xi^R_{tour}}} &=&c_o h H^{-1}_-
 e^{-\beta E} {\bar{Q}}^\dag \ket{\bol{\Gamma}}.
\end{eqnarray}
We can bound:
\begin{equation}
\parallel \ket{\bol{\xi^R_{state}}}\parallel^2 = (c_o h)^2   \; \braok{\bol{\Gamma}}{{\bar{Q}} e^{-\beta E}
H^{-1\dag}_+ H^{-1}_+ e^{-\beta E} {\bar{Q}}^\dag}{\bol{\Gamma}} 
\leq  \lambda_{+\;min}^{-2} \;\; (c_o h)^2 \;  \braok{\bol{  \Gamma}}{\bar{Q} 
 e^{-2\beta E} {\bar{Q}}^\dag }{\bol{ \Gamma}},
\label{bound}
\end{equation}
where $\lambda_{+\;min}$ is the smallest  eigenvalue of $H$ in the two-fermion subspace 
above the gap and
 we have used $\langle \psi | A A^\dag |\psi \rangle \leq |\alpha_{max}|^2
\langle \psi |\psi \rangle$, with $|\alpha_{max}|$ the maximal eigenvalue of $A$.
As  $\braok{\bol{x}} {\bar Q^\dag} {\bf \Gamma}$ is nonzero only in the rotational
region ${\cal D}_\Gamma$ where it can be taken of order one,  in the low temperature limit 
$\braok{\bf\Gamma}{\bar{Q}e^{-2\beta E} {\bar{Q}}^\dag }{\bf\Gamma}^{1/2}
 \sim e^{-\beta E_\Gamma}$
 where $E_\Gamma=min\{E(\bol{x})|\bol{x}\in{\cal D}_\Gamma\}$.
Given that  $\lambda_{+\;min}^{-2}$ is of order one and $c_o$ is
the  normalization of the Gaussian, the contribution above the gap will bounded as
$\parallel
 \ket{\bf \xi^R_{state}}\parallel  \leq c_o  e^{-\beta E_\Gamma} \sim e^{-( E_\Gamma-E_{min})}$
, exactly what we expect of a process that starts in a minimum and climbs up to the region 
${\cal D}_\Gamma$ of energy $E_\Gamma$ where the vorticity is important, and falls 
back again.

The contribution of eigenstates  below the gap is instead given by the loops, 
a fact we shall show in general in the low temperature limit.
The physical meaning of the two-fermion  wavefunctions below the gap 
$\ket{\bol{\chi^\alpha}} \equiv \bar Q \ket{\bol{\xi^{\alpha R}}}$, is now clear 
from equation (\ref{kkk}):
the factor $\braok{\bf \Gamma}{\bar Q}{\bol{\xi^{\alpha R}}}$ will be important 
only if the `vorticity' $\Gamma$ intersects the region where the two-fermion eigenstate 
$\ket{\bol{\chi^\alpha}} \equiv\bar Q \ket{\bol{\xi^{\alpha R}}}$ is non-negligible. Hence,
 each two-fermion partner of
a one-fermion eigenstate below the gap defines the region where a vorticity has to be 
applied in order to excite a current through the corresponding loop.
In the low temperature case, one expects the current through the saddle to be of order 
$e^{\beta (E_{saddle}-E_{min})}$. If $E_\Gamma>E_{saddle}$ the contribution of the loops
 dominate, and the bound above means that it can only be given by
the eigenstates below the gap. Since in simple systems there are only a few of those, the
distribution will not change dramatically with small changes of the vorticity location.

In the general case of systems with a gap in timescale, but with nonzero temperature,
one can still consider forces whose vorticity is `near' or `far' from a state,
and computing the currents induced one can take the construction as a basis for 
 a definition of `loop'. An interesting question is to analyze the effect these loops  have
in the series development of the free energy, an analysis \`a la Langer
\cite{Langer67} would clarify the issue.

\subsection{Induced currents and holes}
\label{sec:induced}
One of the  cases in which it is interesting to calculate currents is when we apply a
constant electric field and join the ends of the sample. We create thus a
 manifold with a hole inside. 
Up to now we have excluded such situations, and indeed some modifications to
the arguments have to be taken into account.
Consider the simple example of a particle in a one-dimensional ring with coordinate $x$,
with $0 \leq x \leq 2\pi$, and no potential.
Both the Fokker-Planck and supersymmetric operators read:
\beq
H= \frac{1}{T}(Q+{\bar{Q}})^2= -T\frac{\partial^2}{\partial x^2}=H_{FP},
\eeq
with 
\beq
Q=-\mathrm{i}T\frac{\partial}{\partial x}
a \;\;\; ; \;\;\; {\bar{Q}}=-\mathrm{i}T\frac{\partial}{\partial x}
a^\dag.
\eeq
The zero fermion states are $\propto e^{ikx}$, with $k$ any integer. In particular, the 
zero-fermion ground state is a constant, $\psi^{0R}(x) =\frac{1}{2\pi}$ as expected.
On the other hand, one-fermion states are also of the form
 $\propto e^{ikx}a^\dag|-\rangle$: we find that we have a one-fermion eigenstate with zero eigenvalue 
$\ket{\bol{\xi^{0R}}}=\frac{1}{2\pi} a^\dag\vidk$, a possibility that we had excluded 
for spaces without holes (see Appendix \ref{qrel}). 
Furthermore, the one-fermion ground state has no partner: 
$\ket{\bol{\xi^{0R}}} \neq  {\bar{Q}}\ket{\bol{\psi^R}} $.
The situation changes when we add a constant field ${\cal{E}}$, so that now:
\beq
H= \frac{1}{T}(Q+{\bar{Q}_{\cal{E}} })^2=-\frac{\partial}{\partial x}(
 T\frac{\partial}{\partial x}+{\cal{E}} )=H_{FP},
\eeq
with 
\beq
Q=-\mathrm{i}T\frac{\partial}{\partial x}
a \;\;\; ; \;\;\; {\bar{Q}}_{\cal{E}}=-\mathrm{i}(T\frac{\partial}{\partial x}+{\cal{E}})
a^\dag.
\eeq
The eigenstates do not change, and we still have the same one and zero-fermion ground
states, but now, remarkably:
\beq
-\mathrm{i}T{\cal{E}}\ket{\bol{\xi^{0R}}}  = {\bar{Q}}_{\cal{E}}\ket{\bol{\psi^{0R}}} \neq 0,
\eeq
so that one and zero fermion ground states have become partners. We also conclude 
that the meaning of the one-fermion ground state is to give the stationary current distribution
around the ring.
A last point to see in this simple example is that when the field is on, the force does not
derive globally from a potential ($E={\cal{E}}x$ would be multiply valued), and we cannot
change globally to the Hermitian basis! 

 Consider in general diffusion in a space with a hole, so that we can have a force
field $\bol{f}$  with everywhere
$\frac{\partial f_j}{\partial x_i}=\frac{\partial f_i}{\partial x_j}$ but {\em not}
deriving from a global potential. We consider a small perturbation 
\begin{eqnarray}
H^f_{FP} &=&-\sum_i\frac{\partial}{\partial x_i} \left(T\frac{\partial}{\partial x_i}+E_{,i}+h f_i\right),
\label{fokkerp}
\end{eqnarray} 
with $h$ a small parameter.
Proposing as  before  a stationary distribution of the form:
$P_{st}=c_o e^{-\beta E}+h P^1$
the current in the stationary state is again given by (\ref{curre}).
It is easy to see that 
the corresponding one-fermion state $\ket{\bol{\xi^R}}= 
\int d^N\!\!x \; \sum_iJ^{st}_i(\bol{x}) a^\dag_i \ket{\bol{x}}\otimes \vidk $ now satisfies:
\beq
Q \ket{\bol{\xi^R}} = 0, \quad \bar Q \ket{\bol{\xi^R}}=0.
\eeq
Also:
\beq
\ket{\bol{\xi^R}} = {\bar{Q}}_{f} \ket{\bol{P_{st}}} =-\mathrm{i} \left[T\frac{\partial}{\partial x_i}+E_{,i}+h f_i\right]\ket{\bol{P_{st}}}=
 -\mathrm{i}h e^{-\beta E} \sum_i  
a^\dag_i\left[c_o f_i + T \frac{\partial \left(e^{\beta E}P_1\right) }{\partial x_i} \right]\vidk.
\label{kkjj}
\eeq
 Just as in the previous example, we have shown that it is the 
nonconservative field that makes $\ket{ P_{st}}$ and $\ket{\bol{ \xi^R}}$ 
become partners (otherwise the last of (\ref{kkjj}) is empty),
and the physical interpretation is that each zero eigenvalue one-fermion eigenstate
corresponds to a current induced around a hole by small fields.
Again, for the perturbed Hamiltonian $H^f_{FP}$ one cannot construct a global Hermitian 
basis as for $H_{FP}$.

Let us conclude this section  with an alternative  variational interpretation for the loops. 
Suppose we ask  which is the field $\bol{f}$ such that it maximizes the  power 
$W=\int d^N\!\!x \;\sum_i f_i J^{st}_i$
done on the system, while having the minimal Gibbs expectation $V$ for its violation
of detailed balance (the `vorticity): 
\begin{equation}
V \equiv \frac{
\int d^N\!\!x\;  e^{-\beta E} \sum_{jk} (\frac{\partial f_j}
{\partial x_k}-\frac{\partial f_k}{\partial x_j})^2
}
{\int d^N\!\!x\;  e^{-\beta E}}.
\eeq
A simple calculation using (\ref{annihilators}) yields:
\beq
{\cal{F}}=\frac{V}{W} \propto
\frac{\sum_\alpha c_\alpha^2 \lambda_\alpha}{\sum_\alpha c_\alpha^2},
\label{calf}
\eeq
where $c_\alpha=\int d^N\!\!x\;\sum_k f_k\xi_k^{\alpha R}$ and
 $\ket{\bol{\xi^{\alpha R}}}$ are 
1-fermion eigenstates annihilated by $Q$ (the 'loops').
In conclusion ${\cal{F}}$ is minimized if ${\bf f}$  is a left, 1-fermion eigenstate
'below the gap'. 
Clearly, the definition is valid at arbitrary temperatures.

\section{The big picture}
\label{sec:thebig}
\subsection{Low temperature structures}
\label{sec:lowt}
At low temperatures eigenstates peak on  structures with dimensions smaller
than $N$, and fall
of exponentially away from them, in a width  that vanishes with $T$.
Right eigenstates `below the gap' are made of linear combinations
of functions peaked on the following structures:

\begin{itemize}

\item 
{\em Zero fermions}: points, the local minima.

\item
{\em One fermion}: gradient lines joining minima through saddle points.
The eigenstates with a zero-fermion partner are the true, open paths, and
those with two-fermion partners are closed loops.

\item
{\em Two fermions}: Two-dimensional surfaces containing a saddle with two unstable 
directions, spanned by all the descending gradient lines emanating from it.  
The eigenstates with a one-fermion partner are peaked on open surfaces (surfaces with borders),
 and those with three-fermion partners are peaked on closed surfaces (borderless surfaces).
 
The physical property of the open surfaces is that a weak nonconservative field $\bol{f}$ 
will generate a current turning around their border, and this  only if the `vorticity' region
in which $\left(\frac{\partial f_j}{\partial x_i} -\frac{\partial f_i}{\partial x_j}\right) \neq 0$ intersects them.

\item
{\em $k$ fermions}: $k$-dimensional surfaces containing a saddle of index $k$,
spanned by all gradient lines descending from it (the unstable manifold of the saddle).

Again, the eigenstates with $k-1$ fermion partners  are peaked on open, and
those with $k+1$-fermion partners on closed surfaces. 
The border of the surface associated with the former is the region where
the $k-1$ fermion partners are peaked.

\end{itemize}

Left eigenstates `below the gap' can be obtained using the fact that a left eigenstate
with $k$ fermions is a {\em right} $N-k$ fermion eigenstate 
of the problem with the inverted potential $-E$ (cfr. Eq. (\ref{gogo})).

One thus 
concludes that $k$ fermion left eigenstates 
 are made of linear combinations
of functions  peaked on the following structures:

\begin{itemize}

\item 
{\em Zero fermions}: constant within a basin of attraction of each minimum (a well known fact).

\item
{\em One fermion}: $N-1$ dimensional basin of attractions of saddles (themselves subsets 
of the borders between basins).

\item
{\em Two fermions}: $N-2$ dimensional 
basins of attraction of saddles with two unstable directions.

\item
{\em $k$ fermions}: $N-k$-dimensional surfaces spanned by the set of descending paths
terminating in each saddle of index $k$ (i.e., the basins of attractions
of these saddles).

\end{itemize}
This is the structure of basins within basins that was argued is relevant in systems
with slow dynamics \cite{Kurchan96}.
One can again distinguish open and closed surfaces, and this is related to
whether the wavefunction of $k$ fermions has a partner with $k-1$ or $k+1$ fermions.

Each of the right eigenstates below the gap are peaked on unstable manifolds 
of the  corresponding critical point, while the left eigenstates are peaked 
on the stable manifolds. In this frame, the $\bar Q$ and $Q$ operators act 
as 'boundary operators' generating what is called in the mathematical 
literature the Morse (co)homology.
We shall derive these results constructively  in the next section.

A simple, three-dimensional  example will illustrate this. Consider
the energy $E(x,y,z)=e(x)+e(y)+e(z)$, where $e(x_i)$ is the symmetric double-well
potential of Fig.  \ref{exemple}. The landscape is defined by a cube with minima in its 
$8$ vertexes, $12$ saddles of index $1$ midway along its sides, $6$ saddles of index $2$ 
at the face centers and one saddle of index $3$ at the cube's center.
The right eigenstates below the gap are as follows {\em i)} $8$  with zero fermions  
located at the vertices's,  {\em ii)} $12$ with one fermion located on the sides, 
of which six are passages and six are loops along the perimeter of the faces,
 {\em iii)} $6$ with two fermions  peaked on the faces, of which five are independent 
open faces and one is the total (closed) surface of the cube, and  {\em iv)} one with 
three fermions constant inside all the interior of the cube.

\subsection{Defining `free energy' structures}
\label{sec:def}
The zero-temperature limit allows us to see very clearly the different structures
that emerge. However, the main point of this paper is that these can be transferred
to a more general situation, provided that there is timescale separation -- whatever its origin.
As we have seen already, the role of local minima in the low-temperature situation
is taken by metastable states, and the role of gradient lines by reaction current
distributions.
In the previous section we also attempted a general definition of `reaction loop', on the basis
of the currents that can be induced by a non-conservative weak force.
One has the possibility
of also defining higher structures associated with fermion subspaces of higher fermion numbers,
 (like borders, basins etc.) in general.
In problems in which one can define a free-energy landscape in a precise way 
(typically mean-fieldish situations), the structures we have defined should recover their
 geometric appearance: states becoming points, reaction distributions becoming lines, etc,
but now in the free-energy landscape, in which 
each point stands for many configurations.
The importance of the construction we have been describing is that it does not rely on such
a landscape: the only assumption is timescale separation.
Once this is given, the Morse-theory constraints on the objects follow automatically,
thus showing that the construction makes geometric sense.
 We shall also see in what follows how these structures are be approached by higher forms
of stochastic equations.

\section{Diffusion dynamics}
\label{sec:diffusion}

\subsection{Dynamics}

In order to obtain the currents we  must find eigenstates  of the supersymmetric 
Hamiltonian (\ref{hsusy}) `below the gap' in the 1-fermion sector. 
This may be achieved by solving (\ref{evol11}) at times larger than the microscopic 
times starting from several initial configurations.
In the zero-fermion case, one in fact simulates the Langevin dynamics (\ref{Lang}) rather 
than solving the Fokker-Plank equation, in order  to obtain metastable states.
The question naturally arises as to which is the diffusion equation that reproduces
(\ref{evol11}) in the one fermion  subspace, and more generally in any $K$-fermion 
subspace
\cite{endnote7}.

Let us do this for one-fermion wavefunctions first.

Consider first one particle carrying an $N$-component  vector degree of  
freedom $\bol{u}$.
Let the position of the particle evolve  as a Langevin process (\ref{Lang}),
and the vector $\bol{u}$ as:
\beq
{\dot{u}}_i= - \sum_j E_{ij}({\bol{x}}) u_j.
\label{u1} 
\eeq
From (\ref{Lang}) and (\ref{u1}), we have that 
the joint distribution function ${\tilde F}({\bol{x}},{\bol{u}},t)$ evolves then as:
\beq
\frac{\partial {\tilde F}({\bol{x}},\bol{u},t)}{\partial t}=
\left[ -H_{FP}+ \sum_{ij}\frac{\partial}{\partial u_i} E_{,ij}u_j \right] {\tilde F}({\bol{x}},\bol{u},t).
\label{uu1}
\eeq
Consider now the evolution of the partial averages:
\beq
R_a({\bol{x}},t) = \int d^N\!\!u \; u_a \;  F(\bol{x},\bol{u},t),
\eeq
\begin{eqnarray}
\frac{\partial R_a({\bol{x}},t)}{\partial t}= \int d^N\!\!u \; u_a \; \frac{\partial {\tilde F}(\bol{x},\bol{u},t)}{\partial t} = \int d^N\!\!u\; u_a \left[ -H_{FP}+ \sum_{ij} \frac{\partial}{\partial u_i} E_{,ij}u_j \right] {\tilde F}({\bol{x}},\bol{u},t) \nn = - H_{FP}  R_a({\bol{x}},t) - \sum_j  E_{aj} R_j({\bol{x}},t),
\end{eqnarray}
where we have integrated by parts on $u_a$. This is equation (\ref{onef}), as announced.

The evolution as in (\ref{u1}) and (\ref{uu1}) has the problem that the
particle may very rarely visit a given region of space, but once it does so have large components
for ${\bol{u}}$. A practical modification is to preserve the norm of the vector attached to the particle.
Putting ${\bol{v}}\equiv {\bol{u}}/|{\bol{u}}|$ we obtain the following equation
for a function  $F({\bol{x}},\bol{v},t)$:
\beq
\frac{\partial  F({\bol{x}},\bol{v},t)}{\partial t}=
\left[ -H_{FP} + \sum_{ij}\frac{\partial}{\partial v_i} \left\{
E_{,ij}v_j -v_i \; \sum_{kl} v_k v_l E_{kl} \right\}
- \sum_{kl} v_k v_l E_{kl}  \right] F({\bol{x}},\bol{v},t).
\label{vv1}
\eeq
Computing the evolution of the partial averages
\beq
R_a({\bol{x}},t) = \int_{|{\bol{v}}|^2=1}  d\bol{v} \;  v_a \;  F(\bol{x},\bol{v},t),
\label{average}
\eeq
the result is again (\ref{onef}).
The diffusional process involved is however quite different. The first term in the square bracket 
in (\ref{vv1}) tells us that the dynamics of the particle is still of the Langevin form. 
The second term now gives for the evolution of ${\bol{v}}$:
\beq
{\dot{v}}_i= - \sum_jE_{ij}({\bol{x}}) v_j + v_i  \sum_{kl} v_k v_l E_{kl},
\label{v1} 
\eeq
 which preserves the condition  $\parallel{\bol{v}}\parallel=1$. The third in 
(\ref{vv1}) 
 is a `cloning' term, creating 
  and destroying particles at a rate  $\sum_{kl} v_k v_l E_{kl}$
\cite{endnote8}.
In Fig. \ref{arrows} we show the numerical solution of the diffusion equation
for vector walkers  in a potential.
 \begin{figure}
\centering \includegraphics[angle=0,totalheight=6cm]{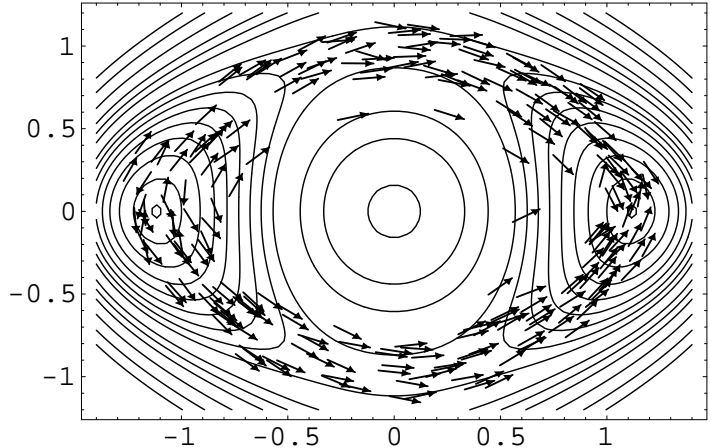}
\caption{Snapshot of a population of walkers in the stationary state. The 
potential is taken from \cite{Dellago98}. 
It has two minima (right and left), two saddles (top and bottom) and a 
maximum in the center. }
\label{arrows}
\end{figure}

\subsection{Low temperature structures}

Let us now use the   equations for ${\bol {v}}$ to show that at low temperatures the 
eigenstates with eigenvalues close to 
zero correspond to lines joining minima through the saddles.
At low temperatures, particles fall along gradient lines:
\beq
{\dot{x}}_i = -E_{,i}(x).
\label{ppp}
\eeq
Assume that the distribution consists of particles
sitting along an isolated  gradient line, and having $v_i=E_i/|\nabla E|$.
Putting this into (\ref{v1}), we find that the  condition $v_i=E_i/|\nabla E|$ 
is preserved as particles fall along the gradient line.
On the other hand, the depletion of particles along the trajectory 
consists of a term due to migration $-\frac{d {\dot{{\bol{x}}_\ell}}}{d \ell}=\frac{
d{\bol{\nabla E}} \cdot {\bol {v}}}{d \ell}$ ($d \ell$ the element length along the line),
and of cloning $\sum_{kl} v_k v_l E_{kl}$: both terms cancel since $d\ell$ is parallel to 
$\bol{v}$. Hence, a uniform distribution of particles along a gradient line
with $v_i=E_i/|\nabla E|$ is stable, provided nothing happens at the ends.
Now, the only possibility for the ends not to destroy stability is that they are stationary 
points, so that there is no particle exchange there. 
 Furthermore, 
 the distribution has to be in particular  peaked along a path joining two
minima through a saddle of order one. The reason is as follows:
particles are constantly falling, 
the measure is preserved because there is a high birth rate  near
the saddle. Now, if the saddle in question is of index higher than one, the slightest noise 
will make particles that are born near it emigrate in other directions, as there is more than 
a single descent path in that case,  rendering the solution along a single gradient line 
unstable. A stable one-fermion solution on a higher dimensional surface
is on the other hand impossible because the surface expansion rate is no longer
compensated by the cloning.

In Appendix \ref{difff}
we give the generalization of the  evolution equations for
higher fermion numbers, and we use them to generalize this argument to show the result announced 
in the previous section 
that low-temperature 
$k$-fermion eigenvalues below the gap are linear combinations of constant densities 
filling  the surfaces spanned by all the descending paths emanating from saddles
of order $k$. The argument is entirely similar to the one for the one-fermion sector:
each particle has a $k$-form attached to it, whose distribution
is preserved as it falls down. The cloning term precisely compensates the effect of
the redistribution of particles, and the solution for $k$ forms is unstable unless
particles are born near a saddle of index $k$.

\subsection{Long-time evolution of {\bf  v}}

At times longer than the inverse of the lowest one-fermion eigenstate,
one can expect that $F(\bol{x},\bol{v},t)$ will converge to a stationary 
distribution symmetric in $\bol{v}$:
\beq
\lim_{t\rightarrow \infty}F(\bol{x},\bol{v},t)=F^{eq}(\bol{x},\bol{v}), \quad 
 F^{eq}(\bol{x},\bol{v})=F^{eq}(\bol{x},-\bol{v}) \quad \forall \bol{v},
\eeq
and the averages  (\ref{average}) vanish
\beq
R^{eq}_a({\bol{x}},t) = \int_{|{\bol{v}}|^2=1}  d^N\!\!v \; v_a \;  F^{eq}(\bol{x},\bol{v},t)=0.
\eeq
To compensate this mean death rate, the usual practice in Diffusion Monte Carlo
schemes is to add an overall cloning probability (see \cite{endnote8}).
In any case,  one  can show that $F^{eq}(\bol{x},\bol{v})$ itself will be 
peaked paths at low temperatures.
To do this, it suffices remake the argument of Section \ref{sec:thebig} A. 

\section{Path sampling: Lyapunov weights}
\label{sec:path}
The same ideas can be written in the path-integral formalism.
Let us start by computing
\bea
I(\bol{x_0},\bol{x_1})=
 \sum_i \vidb \otimes \braok{\bol{x_0}}{a_i \; e^{-H t}
 \; a^\dag_i}{\bol{ x_1}}\otimes \vidk= 
\nn
 \sum_{\alpha i} 
\xi^{\alpha L}_i(\bol{x_1})\xi^{\alpha R}_i(\bol{x_0})
\; e^{-\lambda_\alpha t} 
=\sum_{\alpha i} 
\xi^{\alpha h}_i(\bol{x_1})\xi^{\alpha h}_i(\bol{x_0})
\; e^{-\lambda_\alpha t}.
\eea
Because $H$ is quadratic in the fermions, the evolution for the $a^\dag_i$ is 
linear, and we have in terms of the trajectories  
\cite{Zinn-Justin96,Tanase03a}:
\begin{equation}
I({\bf  x_0,x_1}) =
 \left< \int \; D[paths]  \;  {\mbox{Tr}}{\bf U}_{path} \; \; 
 \Pi_l \; \delta[{\dot{x}}_l+E_{,l}-\eta_l]
 \right>_\eta =  \int D[paths] \;   {\mbox{Tr}}{\bf U}_{path} \;
 e^{-S_{path}},
\label{partition}
\end{equation}
where the sum is over all paths going  from $\bol{x_1}$ to $\bol{x_0}$, and the
 average is over the noise $\eta$ 
realization. $S_{path}$ is the usual Langevin action \cite {Zinn-Justin96}
\beq
S_{path}=\int_0^t d\tau \frac{1}{4T}\sum_i[\dot z_i^2 + E_{,i}^2-2TE_{,ii}]+ 
\frac{1}{2T}[E(\bol{x_1})-E(\bol{x_0})].
\eeq
 ${\bf U}(t)$ is the matrix solution of the linear equation 
\beq
\dot U_{ij}=- \sum_k E_{,ik}U_{kj} \quad \quad U(0)=\bol{I},
\eeq
which depends on the path through $E_{,ik}$. It describes the linear transformation of a small
region around the trajectory, defined by a set of nearby initial conditions {\em and the same
 thermal noise}. 
With these notations, we have:
\beq
I(\bol{x_0},\bol{x_1})=\sum_{\alpha i} 
\xi^{\alpha h}_i(\bol{x_1})\xi^{\alpha h}_i(\bol{x_0})  e^{-\lambda_\alpha t}=  \int D[paths]  \; 
e^{-(S_{path} - L^1_{path})},
\label{weight}
\eeq
where we have defined the (pseudo) Lyapunov exponent (with the time included!), for large $t$:
\beq
L^1_{path} = \ln \left[ {\mbox{Tr}}{\bf U}_{path} \right].
\eeq
The prefix `pseudo' is a reminder of the fact that actually,  true Lyapunov exponents 
is defined on the basis of the trace of ${\bf U U^\dag}$. For $\bol{x_0} = \bol{x_1}$ our 
matrix ${\bf U}$ is symmetric {\em on average} $\langle U_{ij}\rangle =\langle U_{ji} \rangle$
(a consequence of detailed balance), but not along a single trajectory.
We shall return to this point later.
 For large $t$, $L^1_{path}=\ln |\lambda^U_{max}|$ where  $\lambda^U_{max}$ is the 
 eigenvalue of ${\bf U}_{path}$ with the largest real part.
Note that $L^1_{path}$ can be calculated for long times by considering the path and a 
nearby path starting from an initial condition close to $\bol{x_0}$ and evolving with the same 
noise, just as in the computation of an ordinary Lyapunov exponent, or on the basis
of the force required to keep
the distance between paths fixed.

 \begin{figure}
\centering \includegraphics[angle=0,totalheight=6cm]{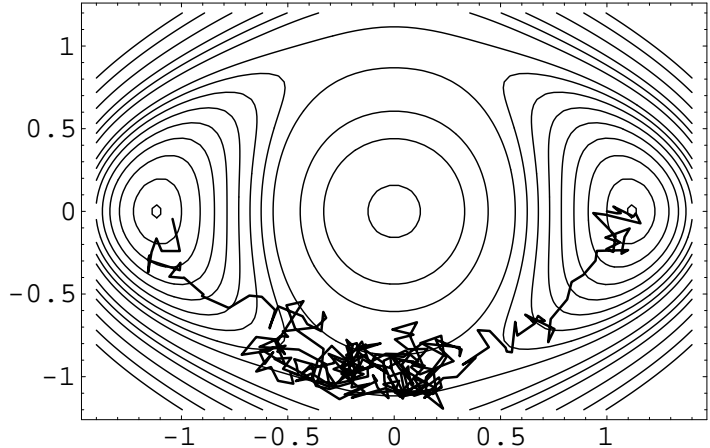}
\caption{One typical trajectory with the ends fixed in two different 
metastable states sampled with Lyapunov+Langevin action. Same potential as in Fig. \ref{arrows}}
\label{chaly}
\end{figure}

Consider the trajectories weighted with the modified action (\ref{weight}).
We wish to know the distribution of $\bol{x}$ at an intermediate time $t'$.
Let us compute the expectation of an arbitrary  function $B(\bol{x})$ at $t'$:
\begin{eqnarray}
\langle B(t',\bol{x_0},\bol{x_1}) \rangle = 
I^{-1}(\bol{x_0},\bol{x_1}) \;\int d^N\!\!x \; \sum_i 
\vidb \otimes \braok{\bol{ x_0}}{a_i e^{-(t-t')H}B(\bol{ x})  e^{-t'H}  
a^\dag_i }{\bol{ x_1}} \otimes \vidk \nn
= I^{-1}(\bol{x_0},\bol{x_1})  \;\int d^N\!\!x \;\sum_{\alpha \beta i} 
\xi^{\alpha L}_i(\bol{x_1})\xi^{\alpha R}_i(\bol{x_0})\;\;
 \braok{\bol{\xi^{\alpha L}}}{B(\bol{x})}
{\bol{\xi^{\beta R}}} e^{-\lambda_\beta (t-t')-\lambda_\alpha t'}.
\end{eqnarray}

Considering $t'$ and $(t-t')$ longer than the microscopic times but much 
shorter than the passage times one can retain only the contributions 'below the gap':
\begin{eqnarray}
\langle B(t',\bol{x_0},\bol{x_1}) \rangle 
= I^{-1}(\bol{x_0},\bol{x_1}) \;\int d^N\!\!x \; \sum_{\alpha \beta i}
\xi^{\alpha L}_i(\bol{x_1})\xi^{\alpha R}_i(\bol{x_0}) \;\;
 \braok{\bol{\xi^{\alpha L}}}{B(\bol{x})}{\bol{\xi^{\beta R}}}.  
\end{eqnarray}
Now, if $\bol{x_0}$ and $\bol{x_1}$ are at the states at the ends of a 
reaction, the factor $ \sum_{i}\xi^{\alpha L}_i(\bol{x_1})\xi^{\alpha R}_i(\bol{x_0})
 = \sum_{i} \xi^{\alpha h}_i(\bol{x_1})\xi^{\alpha h}_i(\bol{x_0}) e^{-\beta E(\bol{x_1})}$
  selects the relevant currents \cite{endnote9}
, and we obtain:
\begin{equation}
\langle B(t',\bol{x_0},\bol{x_1}) \rangle \sim  \;\int d^N\!\!x \;
 \braok{\bol{\xi^L}}{B(\bol{x})}{\bol{ \xi^R }}=   \;\int d^N\!\!x \;
\braok{\bol{\xi^h}}{B(\bol{x})}{\bol{ \xi^h }} = \;\int d^N\!\!x \;
\parallel \bol{\xi^h}(\bol{x})\parallel ^2 B(\bol{x}),
\end{equation}
where we have assumed the reaction is given by a single $\ket{\bol{\xi^R}}$.
What we have shown is that long paths sample the barrier 
$\parallel \bol{\xi^h}(\bol{x})\parallel$. In other words, trajectories have 
ends of the order of the microscopic time in  $\bol{x_1}$ and $\bol{x_0}$, but otherwise spend 
most of their time in the barrier: see Fig. \ref{chaly}. 
  \begin{figure}
\centering \includegraphics[angle=0,totalheight=6cm]{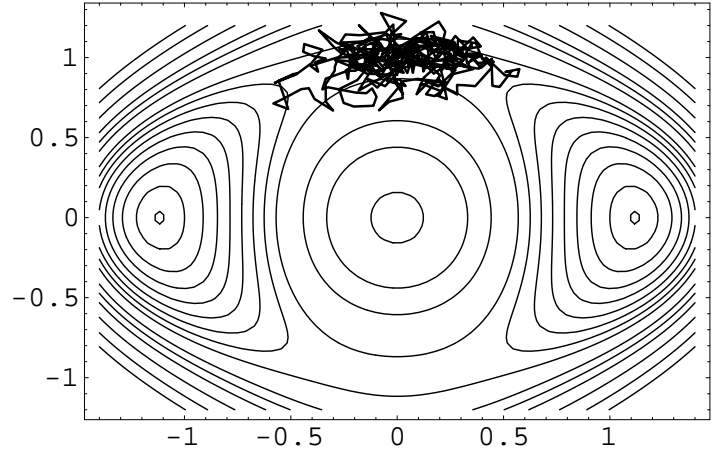}
\caption{One typical closed trajectory sampled with Lyapunov+Langevin 
action. Same potential as in Fig. \ref{arrows}.
}
\label{closed}
\end{figure}
If we consider {\em closed paths} without restrictions on the starting point, we have:
\bea
\int d^N\!\! x_0 \;\langle B(t',\bol{x_0},\bol{x_0}) \rangle = 
\int d^N\!\!x_0 \; d^N\!\!x \; \sum_{\alpha \beta i} 
\xi^{\alpha L}_i(\bol{x_o})\xi^{\alpha R}_i(\bol{x_0})
 \braok{\bol{\xi^{\alpha L}}}{B(\bol{x})}{\bol{\xi^{\beta R}}}
\nn =
\sum_{\alpha \beta } \int d^N\!\! x \; \brak{\bol{ \xi^{\alpha L}}}{\bol{\xi^{\beta R}}}
 \braok{\bol{\xi^{\alpha L}}}{B(\bol{x})}{\bol{\xi^{\beta R}}}
  = \sum_{i\alpha} \int d^N\!\! x \; |\xi^{\alpha h}_i(\bol {x})|^2 B(\bol{x}),
\eea
where the sums are only over the 'eigenvalues below the gap';
this means that we perform a flat sampling over all barriers: see Fig. 
\ref{closed}. If instead of $\parallel \bol{\xi^h}(\bol {x}) \parallel^2$ we wish to 
sample the squared current
$\sum_i \xi^R_i\xi^R_i= e^{-\beta E}\parallel\bol{\xi^h}(\bol {x})\parallel^2$,
 we have to add a Gibbs weight {\em at a single time} in the closed path 
measure.

We can of course fix only one end, and the situation obtained is as in Fig. \ref{ramp} sampling the escape paths from one metastable state.

 \begin{figure}
\centering \includegraphics[angle=0,totalheight=6cm]{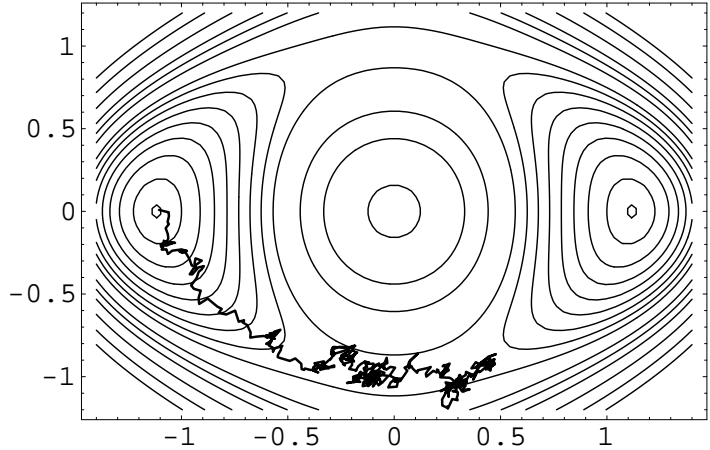}
\caption{ One typical trajectory with one end fixed in a metastable state 
and the other left free, sampled with Lyapunov+Langevin action. 
Same potential as in Fig. \ref{arrows}. }
\label{ramp}
\end{figure}

\subsection{Higher index barriers}

The procedure outlined above can be generalized in a straightforward way to higher
indices. One starts from:
\beq
I^{(k)}(\bol{x_0},\bol{x_1})= \sum_{i_1,\dots,i_k}
 \braok{\bol{x_0}}{a_{i_1}\dots a_{i_k} \; e^{-H t} \; a^\dag_{i_1}\dots a^\dag_{i_k}}{\bol{x_1}},
\eeq
which will select the $k$-fermion eigenstates `below the gap'.
Again, the evolution of each fermion is linear, and a straightforward calculation 
\cite{Gozzi94,Tanase03a} shows that the path-integral reads:
\beq
I^{(k)}(\bol{x_0},\bol{x_1})=  \int D[paths] \; e^{-(S_{path} - L^{(k)}_{path} )},
\label{weightk}
\eeq
where  $L^{(k)}_{path}$ is the   (pseudo) Lyapunov exponent, defined as 
\beq
L^{(k)}_{path}=\ln\left[ \sum_{i_1,\dots,i_k}det_p(U,i_1, \dots,i_k)\right],
\eeq
where $det_p(U,i_1, \dots,i_k)$ are the $k$ minors of U.

For large $t$, $L^{(k)}_{path}$ is the logarithm of the sum of the   $k$ 
eigenvalues $\bol{U}_{path}$ having the largest real parts. 
$L^{(k)}_{path}$ is a measure of the 
expansion of $k$-dimensional surfaces defined by nearby trajectories subjected to the same 
noise \cite{Tanase03a}. Just as before, trajectories weighted with  $L^{(k)}_{path}$ will pile up
in index $k$ barriers.

Before concluding this section, let us point out that, unlike the true Lyapunov
exponents defined in terms of ${\bf U^\dag U}$, the ones we are using here are 
 real  on average, but there could be rare trajectories for which they are imaginary.
The practical procedure is then to separate trajectory space in those that have a real,
 and those that have imaginary
value of  $L^{(k)}_{path}$. As usual in these cases, the separation is natural
since $L^{(k)}_{path}$ diverges in the frontier.

\section{Conclusions}

In this article we have shown how the constructions based on supersymmetry
can shed new light on statistical mechanical questions, providing definitions and
computational schemes for barriers beyond the low-temperature or the mean-field cases.

We have deliberately avoided maximal generality at each step, in various cases 
leaving the most general derivations for the Appendices. We have also not attempted full
mathematical rigor. The aim has been to convince the reader that all developments
are  elementary, though we believe  quite useful.
There are three important omissions:   

\begin{itemize}

\item
Continuous symmetries,  leading to non-isolated saddle points and barriers: 
the subject of degenerate Morse theory \cite{Witten82}.
The formalism adapts itself rather naturally to this case, so we are confident
that the discussion in this direction can be made more complete. 

\item
Dynamics with inertia (Kramers equation). This is important for practical applications,
in which reaction paths have to be found in systems with inertia.

Hamilton's equations do possess a supersymmetry, as shown by Gozzi and Reuter (see \cite{Gozzi94}
and references therein), who used it to rederive 
 some very early results by Ruelle \cite{Ruelle76} where Hamiltonian, 
as opposed to Langevin, dynamics is used to study the topology of the space.
Part of this supersymmetry survives for the Kramers equation  
\cite{TNKfut}, so the results in this paper, and indeed all the construction related to
Morse Theory, can be extended to that case.  
This is a promising line of research, with a considerable number of applications
including, apart from the study of reaction paths, the determination of 
Ruelle-Pollicott  resonances  in chaotic systems.

\item
One can ask if the approach presented in this paper is applicable only to smooth
 energy functions in  space, since there are  real problems with energy functions which 
have singularities: exclusion or Coulomb interactions, hard walls, etc. 
It is possible to  obtain rules for the
behavior of particles (or rather, the forms attached to them)
after a collision with a hard wall without having to integrate
the bounce trajectory every time, by deriving the effect of a regularized 
wall in the limit of infinite steepness.
  Using this method it is possible to construct diffusion equations for
 problems which are entirely entropic, such as hard spheres.

\item
The question of the application to  full quantum evolution remains open. There is of course 
the less general 
possibility of coupling the present scheme to a Carr-Parrinello approach, the dynamics 
being essentially classical in that case.

\end{itemize}

Work is in progress \cite{Tanase03b,TNKfut},
stimulated by the prejudice that things that are pleasant should also be useful.

\vspace*{1cm}
{\bf Acknowledgments.}  We thank A. Oancea for useful discussions.

\appendix
\section{}
\label{qrel}

In his Appendix we sketch the proof that the only zero energy eigenstates of the
 Hamiltonian (\ref{hsusy}) is the Gibbs measure (\ref{rightleft}). 
The proof will be made using the hermitian basis.

Each eigenstate $\ket{\bol{\psi^h}}$ of energy 
zero must be annihilated by both $Q$ and $\bar Q$. 
Indeed, the definition of the zero-energy state:
\bea
0=\braok{\bol{\psi^h}}{H^h}{\bol{\psi^h}}&=&
\braok{\bol{\psi^h}}{\frac{1}{T} ({\bar Q}^h+Q^h)^2}{\bol{\psi^h}}=\nn
\braok{\bol{\psi^h}}{\frac{1}{T} ({\bar Q}^{h\dag}{\bar Q}^h+Q^{h \dag}Q^h)}{\bol{\psi^h}}&=&
\frac{1}{T}(\parallel{\bar Q}^h\ket{\bol{\psi^h}}\parallel^2+\parallel 
Q\ket{\bol{\psi^h}}\parallel^2),
\eea
is equivalent to
\beq
Q^h\ket{\bol{\psi^h}}=0,  \quad  {\bar Q}^h\ket{\bol{\psi^h}}=0. 
\label{annih}
\eeq

Let $\ket{\bol{\psi^h}}$ be a state with $p>0$ fermions such that $\bar Q^h\ket{\bol{\psi^h}}=0$. If we construct the state
\beq
\ket{\bol{\chi^h(y)}}=\mathrm{i} e^{-\beta E(\bol{y})/2}\int^1_0dt\;t^{p-1}\sum_iy_ia_ie^{\beta E(\bol{y}t)/2}\ket{\bol{\psi^h}(\bol{y}t)},
\eeq
one can easily verify that
\beq
\bar Q^h \ket{\bol{\chi^h}}=\ket{\bol{\psi^h}}.
\eeq
This in turn means that
\beq
\braok{\bol{\chi^h}}{{Q}^h }{\bol{\psi^h}}=\braok{\bol{\psi^h}}{{\bar Q}^{h} }{\bol{\chi^h}}=\parallel \ket{\bol{\psi^h}}\parallel^2 > 0,
\eeq
which is incompatible with $Q^h\ket{\bol{\psi^h}}=0$; we have proved that states with 1 
fermion or more cannot be annihilated simultaneously by $\bar Q^h$ and $Q^h$
therefore they cannot have zero energy \cite{Abraham88}.

On the other hand, if $\ket{\bol{\psi^h}}$ has no fermions, (\ref{annih}) can be written as:
\beq
\left(T \frac{\partial}{\partial x_i}+\frac{1}{2}E_{,i}\right) \psi^h(\bol{x})=0 \quad \forall i,
\eeq
and this has only one  solution $\psi^{h}(\bol{x})=c_o e^{-\beta E(\bol{x})/2}$ 
(\ref{rightleft}).

Using standard  arguments one can show that, for an energy $E$ bounded from 
below and satisfying (\ref{integre}), $\ket{\bol{\chi^h}}$ 
and $e^{-\beta E/2}$ have a norm, and thus are in the 
Hilbert space associated with $H^h$. The conclusion is that the only zero 
energy state of  $H^h$ is
\beq
\ket{\bol{\psi^{0h}}}=e^{-\beta E/2}\otimes|->.
\eeq

\section{}
\label{exotic}

In this appendix we write the Morse inequalities derived in section \ref{sec:morse} 
for the trivial topology in a  more general context.
The number of exact zero energy states,
for each fermion sector (let us call them $B_p$) does not depend on the energy 
but only on the topology of the space. To see  this, suppose that the energy is changed
by $E({\bf x})\rightarrow E({\bf x})+ \delta E({\bf x})$. A short computation yields,
to first order:
\begin{equation}
\delta H^h = -\frac{1}{2T} \left[ \sum_i \delta E_{,i} a_i\; , \; {\bar Q^h} \right]_+
+  \frac{1}{2T} \left[ \sum_i \delta E_{,i} a_i^\dag \; , \; { Q^h} \right]_+,
\eeq
and this has zero matrix elements between states with zero eigenvalue, as they are
 annihilated by the charges. First order perturbation theory tells us then that
the eigenvalues stay zero. One can also exclude the possibility of a non-zero eigenvalue
becoming zero, by applying the previous argument to the reverse perturbation.

 For $R^N$, we have shown that   the  $B_p$ are
\beq
B_0=1, \quad B_1=0, \quad \dots, \quad B_N=0.
\eeq

Thus, following the same arguments as in section \ref{sec:morse} one can write generalize
the equalities (\ref{morser}) to:
\bea
M_0&=&B_0+K_1,\nn
M_1&=&B_1+K_1+K_2,\nn
&\vdots &\nn
M_N&=&B_N+K_N.
\eea
where, again,  $K_i>0, \forall i$.

\section{}
\label{difff}

The evolution equation (\ref{evol11})
\begin{equation}
\frac{d {\bol{\psi}}}{dt} = -H \; {\bol{\psi}}, 
\label{evol111}
\end{equation}
for a vector 
\beq
{\bol{\psi}}=\sum_{i_1,\dots,i_k} \psi_{i_1,\dots,i_k} a^\dag_{i_1}...a^\dag_{i_k}|- \rangle,
\eeq
reads, in components:
\beq
 {\dot{\psi}}_{i_1<\dots<i_k}=
\sum_\sigma \sum_\alpha \; (-1)^{n(\sigma,\alpha)} \; E_{\sigma(i_1),\alpha} \;\;
\psi_{\sigma(i_2),\dots,\alpha,\dots,\sigma(i_k)},
\label{evolv}
\eeq
where $\sigma$ denotes all permutations of $k$ indices,
 and $n(\sigma,\alpha)$ is the sign of the 
permutation $(i_1,i_2,\dots,\alpha,\dots,i_k)\rightarrow
(\alpha,\sigma(i_1),\sigma(i_2),\dots,\sigma(i_k))$.
The $\psi_{\sigma(i_1),\dots,\sigma(i_k)}$ are antisymmetric with respect to permutations
of indices.
Proposing the evolution  for  functions of ${\tilde {F}}(\bol{x},\bol{u},t)$, where 
$\bol{ u}$ is the set $u_{i_1,\dots,i_k}$, themselves completely antisymmetric:
\beq
\frac{d{\tilde F}}{dt} = - \left[ H_{FP} -
 \sum_{i_1,\dots,i_k} \frac{\partial}{\partial u_{i_1,\dots,i_k} }
\sum_\sigma \sum_\alpha
(-1)^{n(\sigma,\alpha)}\; E_{\sigma(i_1),\alpha} \; u_{\sigma(i_2),\dots,\alpha,\dots,\sigma(i_k)}
 \; \right]{\tilde {F}}(\bol{ x},\bol{u},t),
\label{horror}
\eeq
we can check integrating
by parts that $\psi_{i_1,\dots,i_k}(\bol{x}) = \int d^N\!\!u \; u_{i_1,\dots,i_k} \; 
{\tilde {F}}(\bol{x},\bol{u},t)$
evolves according to (\ref{evolv}).
This in turn means that $\bol{x}$ evolves according to the Langevin equation, while 
\beq
{\dot{u}}_{i_1,\dots,i_k}= -
\sum_\sigma \sum_\alpha\; (-1)^{n(\sigma,\alpha)} \; E_{\sigma(i_1),\alpha}
\; u_{\sigma(i_2),\dots,\alpha,\dots,\sigma(i_k)}.
\label{horror1}
\eeq

We can also write an equation for normalized variables:
\beq 
v_{i_1,\dots,i_k} = \frac{ u_{i_1,\dots,i_k}}{\sqrt{\sum_{j_1,\dots,j_k} u_{j_1,\dots,j_k}^2}},
\label{norma}
\eeq
such that (\ref{horror}) becomes:
\begin{eqnarray}
\frac{d F}{dt} &=& - \left[ H_{FP} -
 \sum_{i_1,\dots,i_k} \frac{\partial}{\partial v_{i_1,\dots,i_k} }
\left\{ \sum_\sigma \sum_\alpha
(-1)^{n(\sigma,\alpha)} E_{\sigma(i_1),\alpha} \; v_{\sigma(i_2),\dots,\alpha,\dots,\sigma(i_k)}
-  v_{i_1,\dots,i_k} \; {\cal{N}}(\bol{v}) \right\} \right. \nonumber\\
 &+&  {\cal{N}}(\bol{v})\Bigg] F(\bol{x},\bol{v},t),
\label{horrorr}
\end{eqnarray}
where:
\beq
 {\cal{N}}(\bol{v})= \sum_{i_1,\dots,i_k}   v_{i_1,\dots,i_k}  \sum_\sigma \sum_\alpha
(-1)^{n(\sigma,\alpha)}\; E_{\sigma(i_1),\alpha} \; v_{\sigma(i_2),\dots,\alpha,\dots,\sigma(i_k)}.
\eeq
The particles perform Langevin  diffusion, while the equation of motion for the
${\bol{v}}$ read:
\beq
{\dot{v}}_{i_1,\dots,i_k}= - 
\sum_\sigma \sum_\alpha\; (-1)^{n(\sigma,\alpha)} \; E_{\sigma(i_1),\alpha}
\; v_{\sigma(i_2),\dots,\alpha,\dots,\sigma(i_k)} -  v_{i_1,\dots,i_k} \; {\cal{N}}(\bol{v}) ,
\label{horror123}
\eeq
thus preserving the normalization (\ref{norma}). There is also cloning, proportional
to ${\cal{N}}(\bol{v})$.

The equations of motion for ${\bol{v}}$ have an interesting interpretation.
Consider a point ${\bol x}$ and a small oriented $k$-volume element determined by
\beq
\bol{V^k} \equiv (\bol{x}+\delta \bol{x_1})\wedge (\bol{x}+\delta \bol{x_2})\wedge  \dots 
\wedge (\bol{ x}+  \delta \bol{x_k})={\cal{M}} \sum_{v_{i_1,\dots,i_k}}v_{i_1,\dots,i_k}
{\hat e}_{i_1} \wedge \dots  \wedge {\hat e}_{i_k},
\label{surface}
\eeq
where $\wedge$ is the external (wedge) product and ${\hat e}_1$ are the basis vectors.
We have separated the norm ${\cal{M}}$ from the (normalized) `direction'
$\bol{v}$ of $\bol{V^k}$. It is straightforward to see \cite{Tanase03a} that equation
(\ref{horror123})  indeed gives the evolution of $\bol{ v}$ 
as the points are carried by the drift, and ${\cal{N}}={\cal{\dot{M}}}$ gives the expansion rate
of the norm of $\bol{V}^k$. This property is at the basis of the use
of the present formalism to study Lyapunov exponents \cite{Tanase03a}.

Now we can outline a proof that the right $k$-fermion eigenstates  `below the gap' 
are concentrated on  $k$-dimensional surfaces spanned by the trajectories descending
from a saddle of index $k$.
Let us propose that on such surface we have particles whose $\bol{ v}$ at each point
is tangential to such a surface (i.e. it can be generated as in (\ref{surface})),
and  the density of such particles is constant along the surface.
Both features are preserved by the evolution. First, as the particles go downhill,
their forms change so as to remain tangential: this is because their evolution are
precisely based on the linearized equation on the tangential space.
Secondly, the cloning rate is exactly opposite to the expansion rate of a small volume advected
downhill: as a region expands its population increases and vice-versa.
Furthermore, close to the saddle point the surface density expansion rate is $-A_1,\dots,-A_k$,
($A_i$ the Hessian's eigenvalue) while the cloning rate
for a $p$-form is: $-A_1,\dots,-A_p$. Because by assumption $A_1 < 0,\dots, A_k < 0$ and
$A_{k+1} > 0, \dots , A_N > 0$, the cloning is insufficient to maintain
a stationary situation unless $p=k$.


\begin{thebibliography}{60}
\bibitem{endnote2}{Historically, the mapping between supersymmetric systems and stochastic dynamics has been derived by Nicolai \cite {Nicolai80a, Nicolai80b, Nicolai82} and the relation has been further clarified in \cite {Parisi82, Cecotti83}.}
\bibitem{endnote3}{Using supersymmetry and induction, one can show, although we shall not need the result here, that the zero eigenvalues stay zero to all orders in $T$, and only pick up exponentially small corrections.}
\bibitem{endnote5}{In the small temperature limit $t_2 \gg t_1$ meaning that, even though the equilibration inside the wells can be fast, the appearance of a steady current is much slower.}
\bibitem{endnote6}{This is a global comparison of currents. At these times it could still happen that in regions where the current is negligibly small it is not given by the eigenvectors below the gap. If such a detailed approximation is needed, one has to go to longer times.}
\bibitem{endnote7}{ The question has been addressed earlier by people looking for Nicolai \cite {Nicolai80a, Nicolai80b, Nicolai82} maps for SUSY systems in the $K$-fermion sector (see for example \cite{Graham86} and references therein).}
\bibitem{endnote9}{We fix the end states in order to force the paths to go from one metastable state to the other. One can do the same thing by using, instead of the Dirac-delta states $|{{\protect \boldsymbol  {x}}}\delimiter "526930B $ other functions defining in some way the metastable state, in a manner close to that used in `Transition Path Samplig' \cite {Dellago98, Bolhuis02}.}
\end{thebibliography}
\end{document}